\documentclass[a4paper,fleqn,usenatbib]{mnras}
\usepackage{newtxtext,newtxmath}
\usepackage[T1]{fontenc}
\usepackage{ae,aecompl}

\usepackage[colorinlistoftodos]{todonotes}
\usepackage{soul}

\title[N-body simulations of terrestrial planet growth]{N-body simulations of terrestrial planet growth with resonant dynamical friction}

\author[S. Wallace and T. Quinn]{
Spencer C. Wallace\thanks{E-mail: scw7@uw.edu (SCW), trq@astro.washington.edu (TRQ)}
Thomas R. Quinn\footnotemark[1]
\\
Astronomy Department, University of Washington, Seattle, WA 98195
}

\date{Accepted 2019 August 12. Received 2019 August 12; in original form 2018 October 16}

\pubyear{2019}

\begin{document}
\label{firstpage}
\pagerange{\pageref{firstpage}--\pageref{lastpage}}
\maketitle

\begin{abstract}
We investigate planetesimal accretion via a direct N-body simulation of an annulus at 1 AU orbiting a 1 $M_{\sun}$ star. The planetesimal ring, which initially contains N = $10^6$ bodies is evolved into the oligarchic growth phase. Unlike previous lower resolution studies, we find that the mass distribution of planetesimals develops a bump at intermediate mass after the oligarchs form. This feature marks a boundary between growth modes. The smallest planetesimals are packed tightly enough together to populate mean motion resonances with the oligarchs, which heats the small bodies, enhancing their growth. If we depopulate most of the resonances by decreasing the width of the annulus, this effect becomes weaker. To clearly demonstrate the dynamics driving these growth modes, we also examine the evolution of a planetary embryo embedded in an annulus of collisionless planetesimals. In this case, we find that the resonances push planetesimals away from the embryo, decreasing the surface density of the bodies adjacent to the embryo. This effect only occurs when the annulus is wide enough and the mass resolution of the planetesimals is fine enough to populate the resonances. The bump we observe in the mass distribution resembles the 100 km power law break seen in the size distribution of asteroid belt objects. Although the bump produced in our simulations occurs at a size larger than 100 km, we show that the bump location is sensitive to the initial planetesimal mass, which implies that this feature is potentially useful for constraining planetesimal formation models. 
\end{abstract}


\section{Introduction} \label{sec:intro}

The standard scenario of terrestrial planet formation involves the pairwise accretion of small rocky bodies, called planetesimals, that condense from dust out of the protostellar disc \citep{1969Icar...10..109S}. This accretion process can be broken into a series of distinct stages. First, dust particles settle toward the midplane of the disc and clump together via gravitational instability \citep{1973ApJ...183.1051G, 2002ApJ...580..494Y}, streaming instabilities \citep{2007Natur.448.1022J, 2015SciA....1E0109J}, turbulent concentration \citep{2010Icar..208..505C, 2010Icar..208..518C, 2008ApJ...687.1432C, 2016MNRAS.456.2383H} or direct sticking \citep{2013ApJ...764..146G, 2013A&A...557L...4K, 2012ApJ...752..106O, 2012A&A...540A..73W}. These formation models predict a wide variation in the initial size of planetesimals, ranging from hundreds of meters up to a few hundred kilometers in diameter. 

After this condensation process ends, growth continues via pairwise collisions between planetesimals. Above about 1 km in size, gravitational focusing becomes effective and the collision rate strongly increases. This marks the beginning of the intermediate stage of accretion. During this phase, runaway growth \citep{1989Icar...77..330W, 1996Icar..123..180K, 2009Icar..203..626B} quickly increases the size of the largest bodies. Eventually, the largest bodies grow massive enough to heat the small planetesimals, which then inhibits the runaway growth effect. This marks the transition into the phase of oligarchic growth in which the handful of large bodies all tend to grow at the same rate. This produces a bimodal size distribution of planetesimals, with many small bodies and a handful of very large bodies. During this phase, a combination of scattering events between the oligarchs and dynamical friction from the small planetesimals places the oligarchs on nearly circular orbits that are spaced apart by approximately 5-10 Hill radii \citep{1998Icar..131..171K}. Eventually, the oligarchs accrete most of the available material in the vicinity of their orbit, which eventually throttles their growth rate. On much longer time scales, the large bodies perturb each other onto crossing orbits, forming Mars to Earth sized bodies via occasional collisions. This phase of late stage accretion is highly chaotic and takes much longer to play out than the previous stages \citep{1998Icar..136..304C, 2006Icar..183..265R}.\\

Generally, planetesimal accretion models fail to produce a configuration of planets that resembles that of the Solar System. Planets produced in the vicinity of Mars are systematically too massive \citep{1992Icar..100..307W, 2009Icar..203..644R, 2010Icar..207..517M, 2015MNRAS.453.3619I} and the terrestrial planets that form are too eccentric compared to their Solar System counterparts \citep{1998Icar..136..304C, 1999Icar..142..219A, 2001Icar..152..205C}. In addition, the present day water content of Earth, along with the large D/H ratio of Venus \citep{1982Sci...216..630D} does not match with models of solids that condensed out of the Solar nebula at 1 AU. 

These issues all have proposed solutions, which generally involve altering the initial conditions for late-stage
accretion simulations. The condition of the Solar System at the beginning of the late stage accretion phase is poorly constrained. This is mostly due to the fact that planetesimal formation, and therefore the intermediate accretion phase which produces the planetary embryos, is not well understood. Although the present day population of small Solar System bodies has continued to evolve since the end of the intermediate accretion phase, the current size frequency distribution (SFD) of asteroid belt and Kuiper belt objects contains some clues about the accretion history. \citet{2009Icar..204..558M} argued that the SFD of asteroid belt objects larger than 100 km in diameter has been largely unchanged, aside from a size independent depletion factor. \citet{1989Icar...82..402D} did a long term stability analysis of small Solar System objects and found that small bodies left over from accretion should still be largely unperturbed. For these reasons, it should be possible to connect observables in the Solar System to planetesimal formation theories by modelling only the intermediate stages of accretion.

There are two common ways to model planetesimal growth. A powerful approach is to use statistical methods to track the evolution of large groups of planetesimals. This is known as the particle-in-a-box method \citep{1978Icar...35....1G, 1989Icar...77..330W}. The evolution of growth is followed by tracking planetesimals in discrete bins of mass and semi-major axis. This removes the need to calculate the motion of every individual body and allows very large collections of planetesimals to be followed. Unfortunately, the dynamics that governs the evolution of these bodies does not always naturally emerge with this approach. As an example, \cite{2011Icar..214..671W} found that a careful treatment of three body encounters led to a different prediction for the initial size distribution of planetesimals near the asteroid belt. Additionally, the particle-in-a-box method is not well suited for studying oligarchic growth because the largest mass bins, which dominate the evolution in this stage, contain small numbers of bodies. Non-gravitational effects such as gas drag and fragmentation require extra care to implement self-consistently \citep{2008ASPC..398..225L}, although it has been successfully done \citep{1993Icar..106..190W, 2001Icar..149..235I}. To alleviate some of these issues while still being able to model large populations, a newer hybrid approach, in which large bodies are treated as single entities and planetesimals are treated as statistical ensembles has been developed \citep{1997Icar..128..429W, 2006AJ....131.1837K, 2012AJ....144..119L, 2015Icar..260..368M}.

The most reliable and straightforward approach is to use N-body methods to follow the evolution and growth of the planetesimals \citep{1986ApJ...305..564L}. By tracking the individual motions of bodies, the dynamics governing their evolution naturally emerges, no matter what the distribution of bodies looks like. However, N-body simulations involving collision detection are extremely computationally expensive, which severely limits both the resolution and number of timestpng that can be achieved. This is why there are very few studies of runaway and oligarchic with direct N-body simulations in the literature \citep{1996Icar..123..180K, 1998Icar..131..171K, 2002ApJ...581..666K, 2009Icar..203..626B}. Instead, N-body methods are most commonly used to study late stage accretion, where the self-gravity and collisional evolution of the residual planetesimals is largely unimportant \citep{1998Icar..136..304C, 1999Icar..142..219A,  2001Icar..152..205C, 2006Icar..184...39O, 2009Icar..203..644R}.

To date, we are not aware of any N-body simulations that resolve both the runaway and oligarchic growth phase  with more than about $10^4$ particles. At this resolution, stochasticity likely has a significant influence on dynamical friction and resonances may not be sufficiently resolved. \citet{2006Icar..184...39O} found that insufficient planetesimal resolution during the oligarchic growth phase limits the effectiveness of dynamical friction felt by the oligarchs, producing a population of embryos with unrealistically high eccentricities. This idea has also been applied to planet migration through a disc of planetesimals. \citet{2002MNRAS.334...77C} showed that resonances make a significant contribution to the dynamical friction torque exerted by the disc on the planet. This phenomenon requires the planetesimals to be finely resolved \citep{2007P&SS...55.2121B}. Dynamical friction is also facilitated via resonances in galactic dynamics \citep{1972MNRAS.157....1L}. \citet{2007MNRAS.375..425W, 2007MNRAS.375..460W} examined the effect of N-body particle counts on galaxy dynamics and showed that resonant interactions between the bar and halo were only effective with sufficient particle phase space coverage.

For these reasons, we motivate the need for a high resolution N-body simulation of planetesimal growth. We do this to better understand the effect that dynamical friction has on the intermediate stages of terrestrial planet growth and to examine what predictions a high resolution model makes about the residual population of planetesimals in the Solar System. In particular, resonances which are only effective with fine enough resolution, may have an important influence on planetesimal growth. In this paper, we investigate planetesimal evolution during the runaway and oligarchic growth phases with a direct N-body model. We begin by simulating an annulus of planetesimals with similar resolution to \cite{1998Icar..131..171K} to validate our model. We then run the same configuration with 100x more particles to better understand the effects of resolution on the intermediate stages of terrestrial planet growth.

In section \ref{sec:sim}, we describe the simulation code that we use and provide a detailed description of the collision model. We also summarize the initial conditions used. In section \ref{sec:results}, we present the results of the low and high resolution simulation of terrestrial planet growth and highlight differences between the two. We find that a bump develops in the mass distribution of the high resolution simulation, which does not appear in the low resolution run. This feature manifests itself shortly after oligarchic growth commences and we infer that it is produced by extra heating via mean motion resonances between the oligarchs and small planetesimals. To further demonstrate this effect, we re-run the high resolution simulation with a narrower annulus (depopulating many of the resonances) and show that this reduces the prominence of the bump in the mass distribution. In section \ref{sec:dynfric}, we present a set of collisionless simulations of a planetary embryo embedded in an annulus of planetesimals. This more clearly demonstrates the differences in dynamical behavior between the low and high resolution models. In section \ref{sec:intermed}, we present two more simulations of planetesimal growth at intermediate resolutions and demonstrate that the location of the bump is sensitive to the initial planetesimal mass. Finally, section \ref{sec:discussion} connects these new results to our present understanding of terrestrial planet growth and we discuss the additional stpng necessary to use our results to constrain the initial size of planetesimals in the Solar System.

\section{Simulations} \label{sec:sim}

\subsection{Numerical Methods} \label{sec:numerical}

All of the simulations described in this paper were performed with the highly parallel N-body code {\sc ChaNGa} \footnote{A public version of {\sc ChaNGa} can be downloaded from \url{http://www-hpcc.astro.washington.edu/tools/ChaNGa.html}}. {\sc ChaNGa} is written in the {\sc CHARM++} programming language and has been shown to perform well when using up to half a million processors \citep{2015AphCom..2..1} simultaneously. Gravitational forces are calculated using a modified Barnes-Hut tree algorithm with hexadecapole order expansions of the moments. For all of the simulations described in this paper, a node opening criterion of $\Theta_{BH}$ = 0.7 was used. \citet{2000Icar..143...45R} was able to resolve mean motion resonances in a terrestrial disc using a similar tree code with the same node opening criterion, so we expect that our model should properly handle resonance effects as well. The equations of motion are integrated using a kick-drift-kick leapfrog scheme. For more information about the implementation of {\sc ChaNGa} see \cite{2008IEEEpds...ChaNGa}.

\subsection{Collision Model} \label{sec:collModel}

{\sc ChaNGa} is a smoothed particle hydrodynamics code originally designed for cosmology simulations. In order to use {\sc ChaNGa} to study planetesimal coagulation, we implemented a hard-body collision model that treats particles as solid objects with a fixed radius, rather than as smooth tracers of a fluid with a characteristic softening length. This work was largely based off of the hard body collision implementation in {\sc pkdgrav}, which is described in \citet{1994MNRAS.269..493R} and \citet{2000Icar..143...45R}.

Collisions are predicted at the beginning of each drift step by extrapolating the positions of the particles forward using the velocities calculated during the first kick. For each particle, the closest 64 neighbors are considered in the collision search. Because {\sc ChaNGa} is a tree code, the nearest neighbors of a particle can be retrieved in $\mathcal{O}(N\log{}N)$ time. After extrapolating the positions forward and checking for overlap with any neighbors, the earliest collision time $t_{coll}$ is stored for each particle.

After the prediction phase, particles with $t_{coll}$ less than the time step size $\Delta T$ must have their collisions resolved. For simplicity, all collisions in our simulations result in perfect accretion. \cite{1993Icar..106..190W} showed that excluding the effects of fragmentation and collisional rebound does not qualitatively change the growth modes of the planetesimals. A merger between two particles of mass $m_{1}$ and $m_{2}$ results in a single particle of mass $M = m_{1} + m_{2}$, with the radius set to conserve density. The position and velocity of the resulting particle is set to the centre of mass position and velocity of the colliders at the moment of contact. The resulting merged particle is then drifted to the end of the step. If multiple collisions are predicted during a time step, the earliest collision is resolved first. Because resolving a collision can result in another imminent collision, collisions must be resolved one by one, with a new prediction check being run each time.

As in \citet{1996Icar..123..180K, 1998Icar..131..171K} we accelerate the accretion process by artificially inflating the physical radius of the bodies by a factor of $f$. This technique reduces the accretion time-scale by approximately a factor of $f^{2}$, significantly reducing the number of timestpng that must be integrated. Additionally, inflating the particle radii allows us to use a smaller annulus with less planetesimals. The reason we cannot use an arbitrarily skinny annulus is because the edges tend to expand outward due to the unrealistic boundary conditions, decreasing the surface density. The time-scale for this expansion is set by the two body relaxation time-scale, which scales with $N$. Reducing the accretion time-scale by increasing $f$ allows us to study planetesimal growth with a smaller, less computationally expensive annulus.

We must be careful, however, to choose a value of $f$ that is not too large. The rms eccentricity and inclination of the planetesimal disc grows as gravitational encounters transform energy due to Keplerian shear into random motions. By accelerating the growth rate, we cause the transitions between growth modes to happen early, when the disc is less dynamically excited. This discrepancy is partly compensated for by the fact that our model ignores the effects of gas drag, which would damp the eccentricities and inclinations of the planetesimals. We adopt $f=6$ for our calculations, which reduces the amount of gravitational scattering by less than 10\% of its true value and does not qualitatively change the modes of planetesimal growth \citep{1998Icar..131..171K}.

\subsection{Initial Conditions} \label{sec:ics}

We begin by using our collision model to perform two sets of calculations:

(i) $10^{6}$ equal mass bodies with $m = 1.2 \times 10^{21}$ g

(ii) 4000 equal mass bodies with $m = 3 \times 10^{23}$ g

\noindent As previously mentioned, simulation (ii) is meant to be compared with the results of \citet{1998Icar..131..171K}. This allows us to both validate our collision model and also have a baseline to compare our simulations to. In both cases, the planetesimals are distributed randomly in an annulus centred at 1 AU with a width $\Delta a = 0.085$ AU around a 1 $\mathrm{M_{\odot}}$ star. This annulus width was chosen so that the required particle count is minimized without boundary effects influencing planetesimal growth. The surface mass density of the ring is set to 10 g cm$^{-2}$, which approximately corresponds to the minimum-mass solar nebula model \citep{1981PThPS..70...35H} at 1 AU. In case (i), the eccentricities and inclinations are taken from a Rayleigh distribution \citep{1992Icar...96..107I} with $\langle e^2 \rangle^{1/2} = 2 \langle i^2 \rangle^{1/2} = 4 h/a$, where $h$ is the mutual Hill radius. To match the same eccentricity and inclination dispersion with larger planetesimals in case (ii), we use $\langle e^2 \rangle^{1/2} = 2 \langle i^2 \rangle^{1/2} = 0.635 h/a$. In both simulations, the planetesimals are given an internal density of 2 g cm$^{-3}$. We use fixed timestpng with $\Delta T$ = 0.0025 years and evolve the simulations for 20,000 years. Large timesteps diminish the effectiveness of gravitational focusing, which inhibits the collision rate. We find that the collision rate is converged below $\Delta T$ = 0.0025 years. The integration time was chosen to be comparable to the orbital repulsion and Hill radius growth time-scale \citep{1995Icar..114..247K} so that the effects of oligarchic growth are fully realized by the end of the simulation.

It is worth noting that simulation (i) was a very computationally expensive undertaking. In total, the full 20,000 years of integration required approximately 130,000 CPU hours and over 50 days of wall clock time.

\begin{figure*}
    \includegraphics[width=\textwidth]{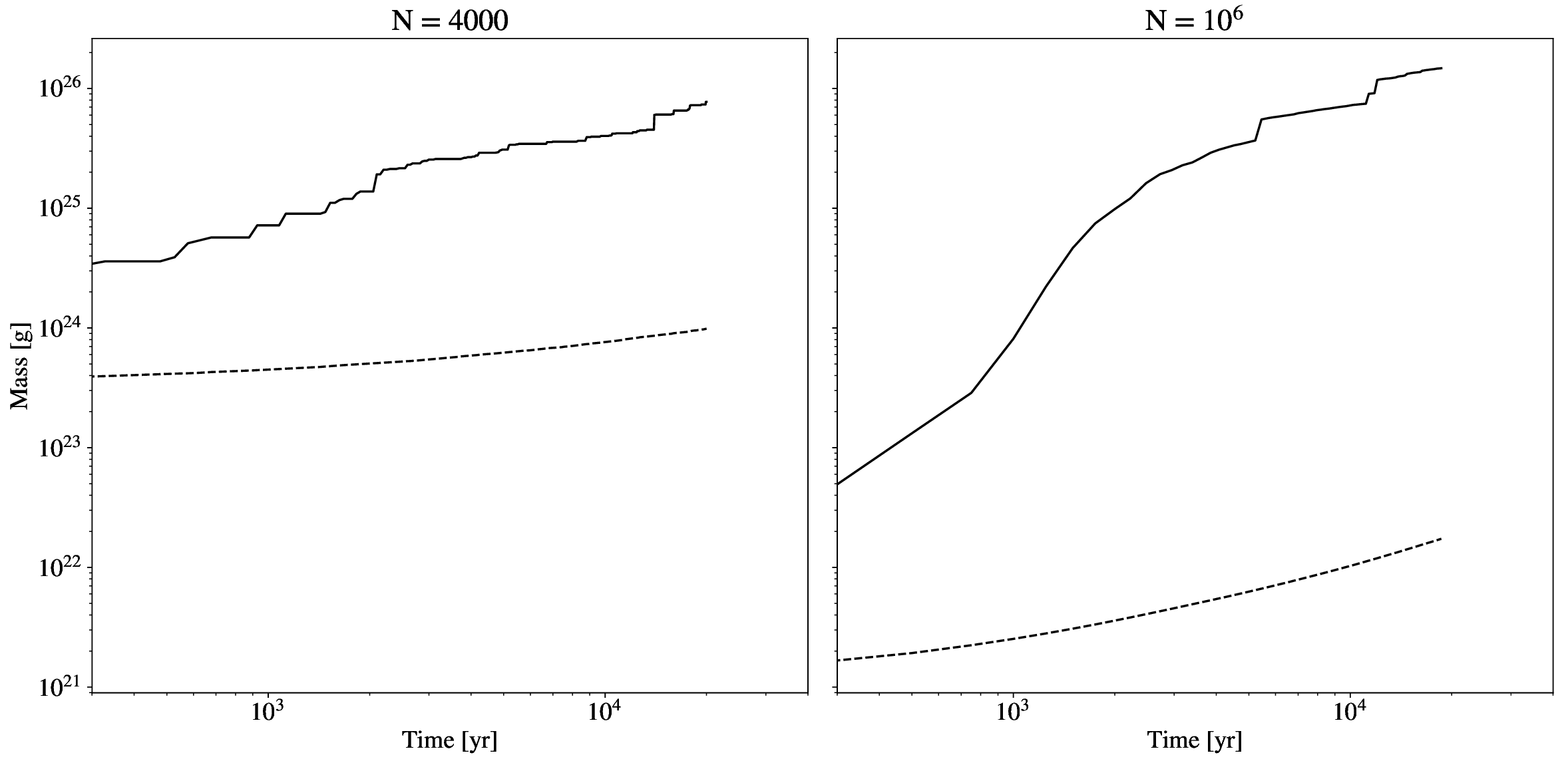}
    \caption{Evolution of the maximum (solid curve) and mean (dashed curve) planetesimal mass in the N=4000 and N=$10^6$ particle simulations. At early times, the maximum mass grows more quickly than the mean mass, which is indicative of runaway growth. After a few thousand years, the separation between the curves becomes a constant factor, signalling the start of oligarchic growth.
    \label{fig:mass_evo}}
\end{figure*}

\begin{figure*}
    \includegraphics[width=\textwidth]{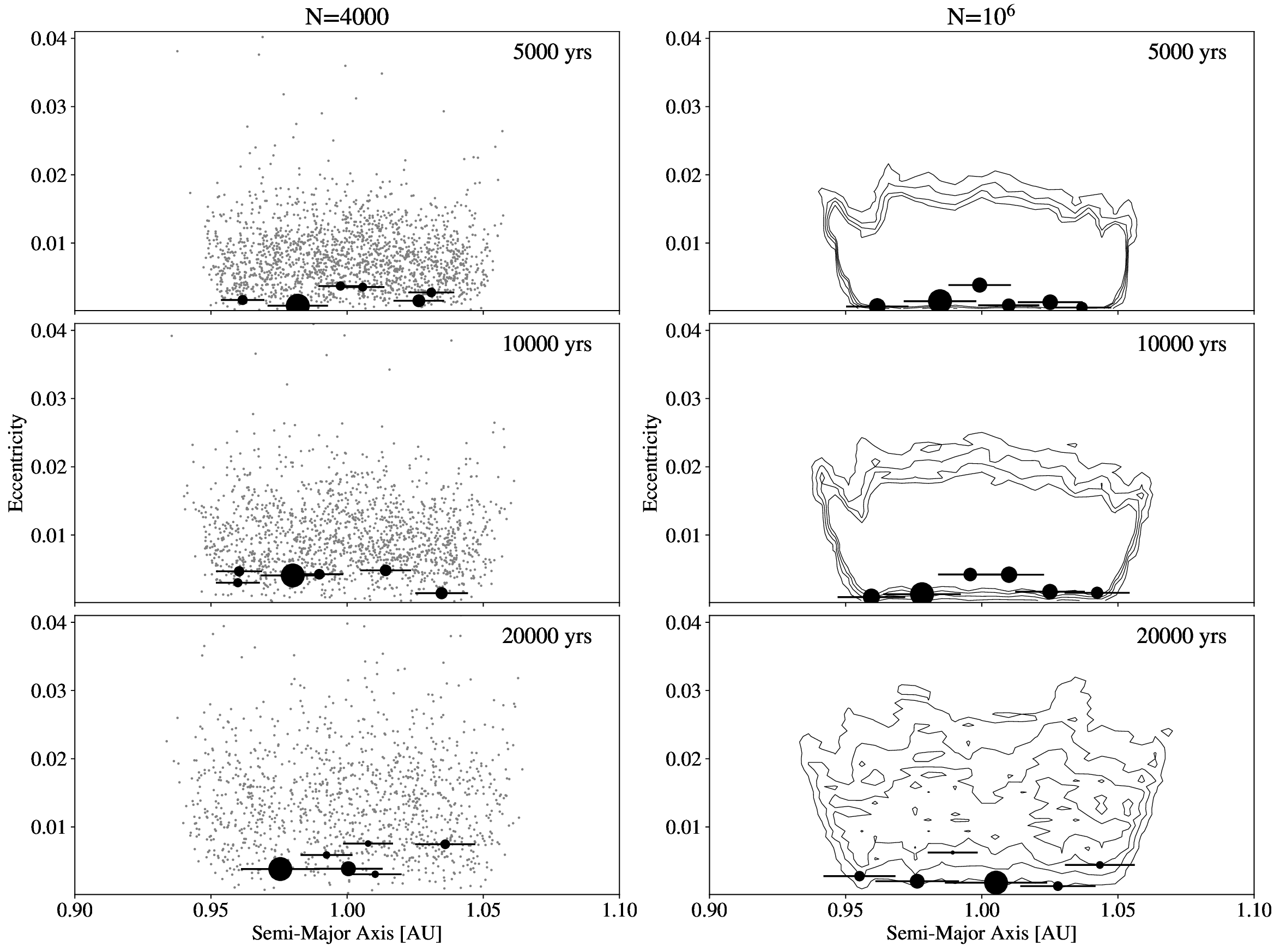}
    \caption{Snapshots from the low and high resolution models in the $a-e$ plane. On the left, the light dots represent individual planetesimals, while the contours on the right hand plots represent curves of constant number density. The contour levels are the same between all panels and correspond to $7.8 \times 10^6$, $1.6 \times 10^7$, $2.3 \times 10^7$, and $3.1 \times 10^7$ planetesimals per AU per unit eccentricity. The black circles denote the configuration of the 6 largest bodies in the simulation, with the area of the circles scaled to the mass of the body. The horizontal error bars are scaled to 5 times the Hill radius of the bodies.
    \label{fig:ae}}
\end{figure*}

\begin{figure*}
    \includegraphics[width=\textwidth]{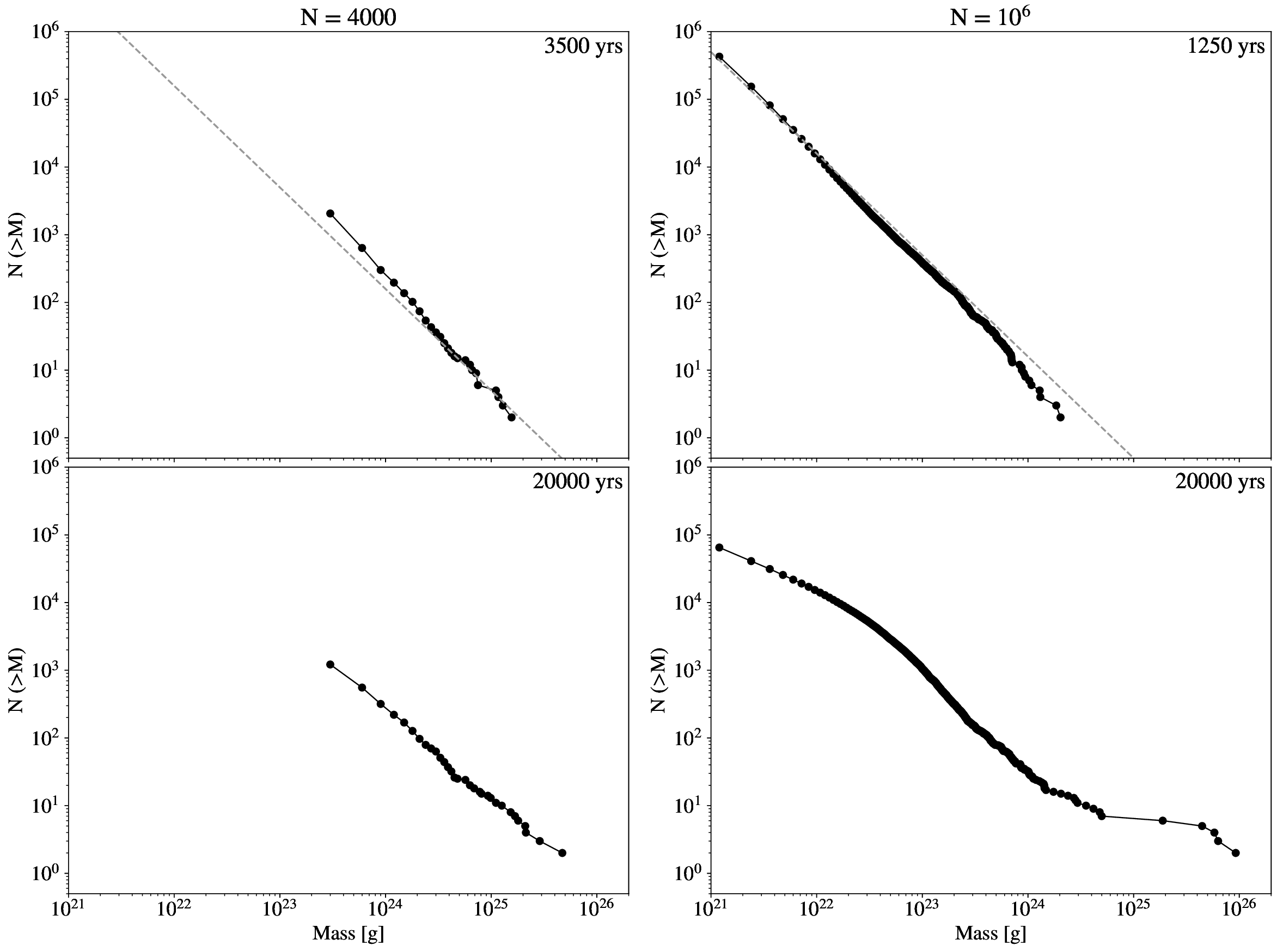}
    \caption{Cumulative number of bodies in each mass bin for the low and high resolution runs, shown at the end of runaway growth (top row) and the end of the simulation (bottom row). The dashed line indicates a slope of -1.5, which is characteristic of runaway growth.
    \label{fig:mass_spectrum_evo}}
\end{figure*}

\section{Results} \label{sec:results}

\subsection{Low vs High Resolution}\label{sec:lowvshigh}

We begin by comparing the evolution of growth between the low resolution (N=4000) and high resolution (N=$10^{6}$) models. As planetesimals collide and grow, gravitational focusing becomes increasingly effective and the relative growth rate increases with mass \citep{1978Icar...35....1G}. Figure \ref{fig:mass_evo} shows the evolution of the average and maximum planetesimal mass in both simulations. Runaway growth at early times is evident from the fact that the maximum mass grows more quickly than the mean mass. Eventually, the largest bodies begin to dynamically heat the neighboring planetesimals, which slows the growth rate of the largest bodies.

When gravitational focusing and dynamical friction are both effective, the growth rate of a planetesimal of mass $M$ is given by

\begin{equation}\label{eq:growth_rate}
\frac{dM}{dt} \propto \Sigma M^{4/3} e_{m}^{-2},
\end{equation}

\noindent where $\Sigma$ is the surface density of solid material and $e_m$ is the rms eccentricity of the planetesimals \citep{1995Icar..114..247K}. Before the oligarchs form, the eccentricity dispersion is independent of mass and the fractional growth rate scales like $dM/dt \propto M^{4/3}$ \citep{1993Icar..106..210I}. This implies that large bodies grow more quickly than small bodies, hence the runaway effect. Once the oligarchs form and dynamical friction becomes effective, energy equipartition causes the velocity dispersion to evolve toward $v_m \propto M^{-1/2}$\citep{1993MNRAS.263..875I}. Using $v_m \propto e_m$ \citep{1993prpl.conf.1061L}, the growth rate during the oligarchic growth phase scales like  $dM/dt \propto M^{2/3}$. Note that this does not imply that smaller bodies grow more quickly than large bodies. Rather, the growth rates tend towards the same value. This is reflected in figure \ref{fig:mass_evo}, where the slope of the maximum mass curve flattens out to match the slope of the mean mass curve after a few thousand years.

Comparing the two panels in figure \ref{fig:mass_evo}, it is also evident that
the rate of growth is more vigorous in the high resolution case. This is due to the fact that the collision time-scale, given by 

\begin{equation}\label{eq:coll_timescale}
t_{coll} = \frac{1}{n \sigma v},
\end{equation}

\noindent is shorter in the latter case, where $n$ is the number density of planetesimals, $\sigma$ is the collision cross section and $v$ is the typical encounter velocity. The collision cross section depends on both the geometric cross section ($\propto N^{-2/3}$) and an extra term due to gravitational focusing ($\propto N^{-7/3}$), where $N$ is the total number of particles in the simulation and the total disk mass is fixed. Because the eccentricity and inclination dispersion (and therefore the scale height of the disk) are kept fixed between the low and high resolution simulations, the typical encounter velocity does not vary with $N$. Here, we retain only the leading order term for the collision cross section, in which case these quantities scale like $n \propto N$, 
$\sigma \propto N^{-2/3}$ and $v \propto const$ so that

\begin{equation}\label{eq:coll_timescale_N}
t_{coll} \propto N^{-1/3}.
\end{equation}

Figure \ref{fig:ae} shows the a-e distribution of planetesimals at three snapshots from both simulations. In both cases, the eccentricity dispersion grows as energy from Keplerian shear is transformed into random motion, an effect known as viscous stirring \citep{2002Icar..155..436O}. The black circles denote the semi-major axis and eccentricity of the 6 largest bodies. The area of the circles indicates the mass of the bodies and the horizontal bars are each scaled to 5 times the Hill radius of the bodies. Gravitational scattering between the oligarchs, coupled with dynamical friction from the surrounding planetesimals, places the oligarchs on low eccentricity orbits that are spaced apart by a 5-10 Hill radii via orbital repulsion \citep{1998Icar..131..171K}.

Next, we examine the evolution of the mass distribution of planetesimals. This is shown in figure \ref{fig:mass_spectrum_evo}. At early times, both models exhibit a power law distribution $N(<M) \propto m^{-p}$, where $p$ = 1.5 for the small bodies. A mass distribution with this slope is characteristic of runaway growth \citep{1993Icar..106..190W}. The top panels show the mass distribution from both simulations at the end of the the runaway growth phase. In all subsequent snapshots, the mass distribution deviates from a single power law as the most massive bodies break away from the distribution, signalling the start of oligarchic growth. This happens sooner in the high resolution case. The bottom panels show the mass distribution at the end of the simulation at T = 20,000 years.

Besides the fact that growth is more vigorous at higher resolution, the final mass distribution of planetesimals in the $N = 10^{6}$ case develops a feature that does not appear in the low resolution model. In the low resolution case, the low mass end of the mass distribution of planetesimals retains a single power law slope. The high resolution model, on the other hand, develops a bump in the mass distribution near $10^{22}$ g.

\begin{figure*}
    \includegraphics[width=\textwidth]{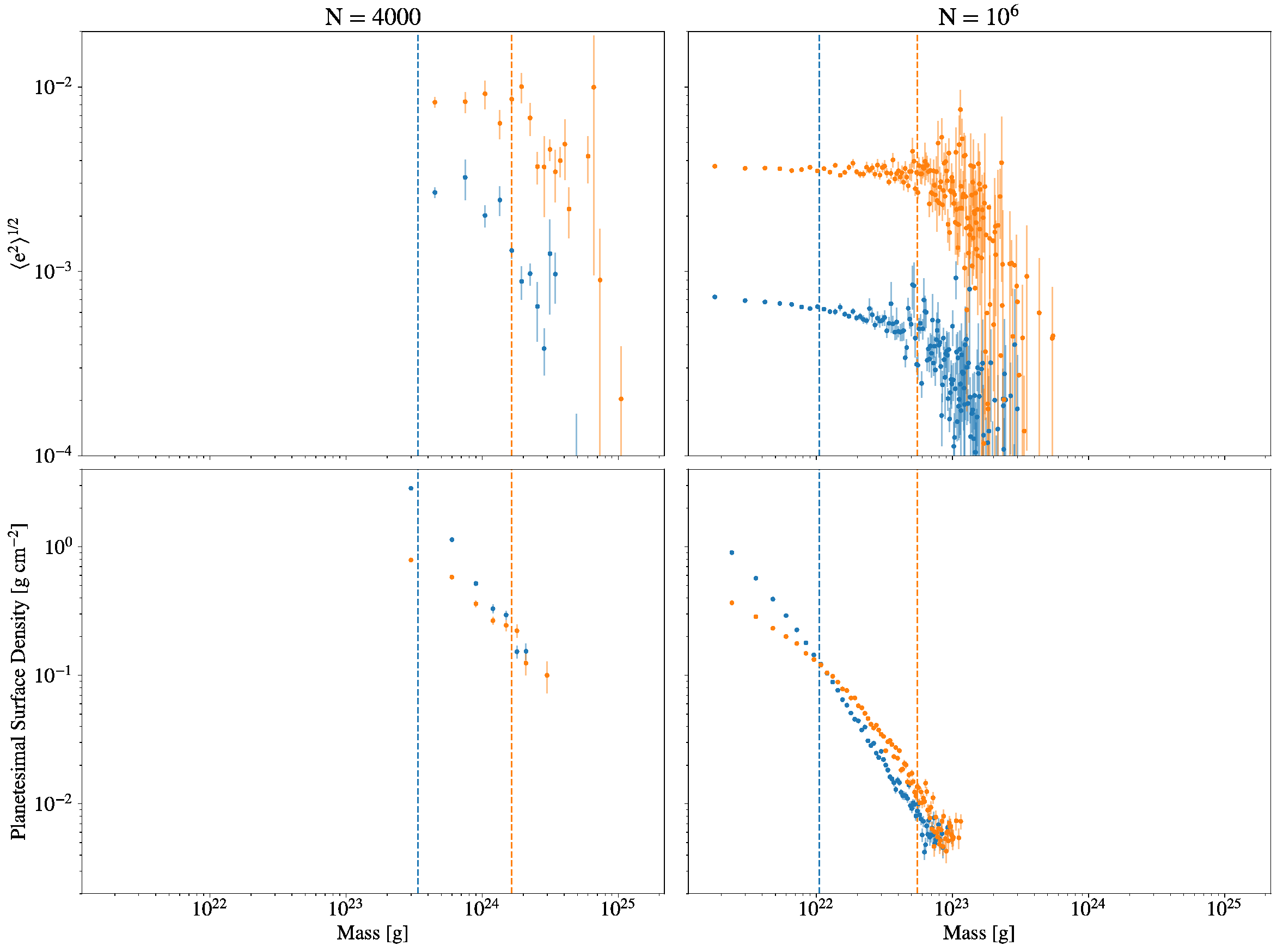}
    \caption{The rms eccentricity (top) and planetesimal surface density at each mass (bottom) shown for the $N = 4000$ and $N = 10^6$ simulations. The blue points correspond to the quantities at the end of the runaway growth phase (when the mass distribution deviates from a single power law) and the orange points are from the end of the simulations (T = 20,000 years). The vertical dashed lines indicate the value of $M_{stir}$ during that snapshot.}
    \label{fig:ecc_den_evo}
\end{figure*}

\subsection{Explaining the Bump in the High Resolution Mass Spectrum}\label{sec:bump}

We would next like to determine what produces the feature at $10^{22}$ g in the high resolution simulation and why it is absent from the low resolution run. Upon close inspection of the simulation snapshots, both models retain a mass distribution that follows a single power law up until the onset of oligarchic growth. The bump in the high resolution mass spectrum becomes visible shortly after the oligarchs form. Over the course of the simulation, the bump gradually becomes more prominent. Because the bump appears shortly after the first oligarchs do, its presence is likely tied to the formation of the oligarchs.

Other investigations of planetesimal growth have revealed a similar bump in the size frequency distribution (SFD) at intermediate masses. \cite{2009Icar..204..558M} found that the location of this bump was set by the initial size of the planetesimals. Objects smaller than this size were created through disruptive collisions, which resulted in a shallower slope on the SFD to the left of the bump. Our collision model does not allow for fragmentation, so we must look for another explanation.

\cite{2011Icar..214..671W} showed that a bump in the SFD was an artefact of the transition between shear and dispersion dominated growth. Objects in the shear regime, whose encounter velocities are dominated by the differential rotation of the disc, grow more slowly than those in the dispersion regime, whose encounters are set by the random velocity dispersion. Because the velocity dispersion varies with planetesimal mass, both of these growth modes can operate simultaneously. Energy equipartition should cause the velocity dispersion to decrease with mass, so the transition between shear and dispersion dominated growth should happen at some intermediate mass. Finally, the smallest bodies have a velocity dispersion that exceeds their escape velocities, which sets a third, slow mode of growth in which gravitational focusing is ineffective. Although our high resolution simulation resolves all three modes of growth, we find that the boundaries between these modes smoothly and steadily evolve. Over the course of the simulation, the shear-dispersion boundary increases from about $10^{22}$ g to $10^{25}$ g, while the dispersion-escape boundary evolves from about $10^{21}$ to $10^{24}$ g. Any artefacts that these boundaries would leave on the planetesimal mass distribution should get washed out, so we also rule these out as explanations for the $10^{22}$ g bump.

Although the boundary between shear and dispersion dominated growth does not seem to be producing the bump, there still must be some kind of transition between growth modes occurring near this mass. As shown by equation \ref{eq:growth_rate}, the growth rate is controlled by both the surface density of the planetesimals and their eccentricities. In figure \ref{fig:ecc_den_evo}, we show the rms eccentricity and surface mass density of the planetesimals as a function of mass at the end of the runaway growth phase (blue points) and at the end of the simulation (orange points). In this figure, each point corresponds to the relevant quantity calculated for all planetesimals with the exact same mass. Because the simulations start with equal mass planetesimals which grow via pairwise collisions, the masses take on discrete, linearly spaced values. We found that logarithmically binning the mass values alters the shape of the distributions, especially in the high resolution case where the quantities span many orders of magnitude. For this reason, we chose not to bin any of the data. The error bars in figure \ref{fig:ecc_den_evo} are obtained via 10,000 iterations of bootstrap resampling. For the high resolution simulation, the error bars at low mass are smaller than the size of the points.

The surface density was determined by calculating an azimuthally averaged density profile using the analysis package {\sc PYNBODY} \citep{2013ascl.soft05002P}. The surface density for each planetesimal mass was taken to be the average surface density for those particles in a single radial bin from 0.9575 to 1.0425 AU, which spans the initial boundaries of the annulus.

One might expect that the rms eccentricity spectrum should eventually reach energy equipartition ($e \propto m^{-1/2}$), but this has been shown to only occur with a sufficiently steep mass distribution \citep{2003AJ....126.2529R}. In our case, the mass distribution is shallow enough that the velocity evolution of the low mass bodies is set only by interactions with large planetesimals. The mass below which this occurs, shown by the vertical dashed lines in figure \ref{fig:ecc_den_evo} is given by \citep{1993Icar..106..210I, 2010Icar..210..507O}

\begin{equation}\label{eq:mstir}
M_{stir} = \frac{\left< m^2 \right>}{\left< m \right>}.
\end{equation}

Below this mass, planetesimals do not produce a dynamical friction wake and their velocity evolution becomes independent of mass \citep{2003AJ....126.2529R}. Because we started with equal mass planetesimals, the mass distribution was steep at early times. A power law slope steeper than $p = -2$ should produce a mass-dependent velocity distribution everywhere \citep{2003AJ....126.2529R}. In the top right panel of figure \ref{fig:ecc_den_evo}, the rms eccentricity distribution is not entirely flat below $M_{stir}$, which is likely due to the evolving mass spectrum. The analysis of \citet{2003AJ....126.2529R}, however, assumes a static mass spectrum.

At late times, a power law break in the surface density distribution forms near $10^{22}$ g in the high resolution simulation. Because the surface density is tied to both the mass distribution and the spatial distribution of planetesimals, it is difficult to learn anything else about the dynamics that are altering the growth rate from this information alone. We will examine this further in section \ref{sec:dynint}.

Given the power law break in the surface density distribution below $10^{22}$ g, we infer that there must be a dynamical mechanism at work that alters the collision rates of the low mass bodies. By the time that the $10^{22}$ g bump begins to appear, the planetesimals are sufficiently hot enough to render gravitational focusing ineffective (equation \ref{eq:growth_rate} is also no longer applicable). In this case, any additional heating actually increases the collision rate. Because this effect appears around the time of the onset of oligarchic growth, it likely has something to do with dynamical friction between the oligarchs and the planetesimals. In the next section, we examine how dynamical friction might be more effective with small planetesimals.

\section{Dynamical Friction and Resolution} \label{sec:dynfric}

Although the Chandrasekhar formula \citep{1943ApJ....97..255C} contains no dependence on particle mass, the 'granularity' of the surrounding medium has been shown to influence the action of dynamical friction \citep{2007P&SS...55.2121B}. This is because the individual kicks from gravitational encounters become less frequent and more powerful at coarse resolution, introducing extra stochasticity as the system evolves toward energy equipartition. \citet{2006Icar..184...39O} showed that a finely resolved planetesimal distribution during the oligarchic growth phase produced planetary embryos with low eccentricities. They did not, however, examine the mass spectrum to see if the oligarchs were preferentially heating the smallest planetesimals. In addition, their simulations only contained a few thousand particles, while our high resolution run contains in excess of 200,000 bodies at the onset of oligarchic growth. Comparing the largest embryos at the end of simulation (i) to their results, we find that the largest embryo in our high resolution run has an eccentricity that is a factor of 2 smaller than the embryos produced by \citet{2006Icar..184...39O}.

In addition to altering the cumulative effect of close gravitational encounters, there is evidence that energy and angular momentum exchange through resonances is more effective with fine granularity. For example, in collisionless simulations of galaxies, \citet{2007MNRAS.375..425W, 2007MNRAS.375..460W} showed that a minimum number of particles was required to populate resonances and couple the rotation of a bar to the central halo cusp through Lindblad resonances. If the resolution was too coarse, gravitational potential fluctuations would scatter particles out of resonances and prevent any strong torque between the bar and the halo. Likewise, \citet{2002MNRAS.334...77C} showed that resonant torque has a measurable effect on the interaction between a planet and a planetesimal disc.

\begin{figure}
    \includegraphics[width=\columnwidth]{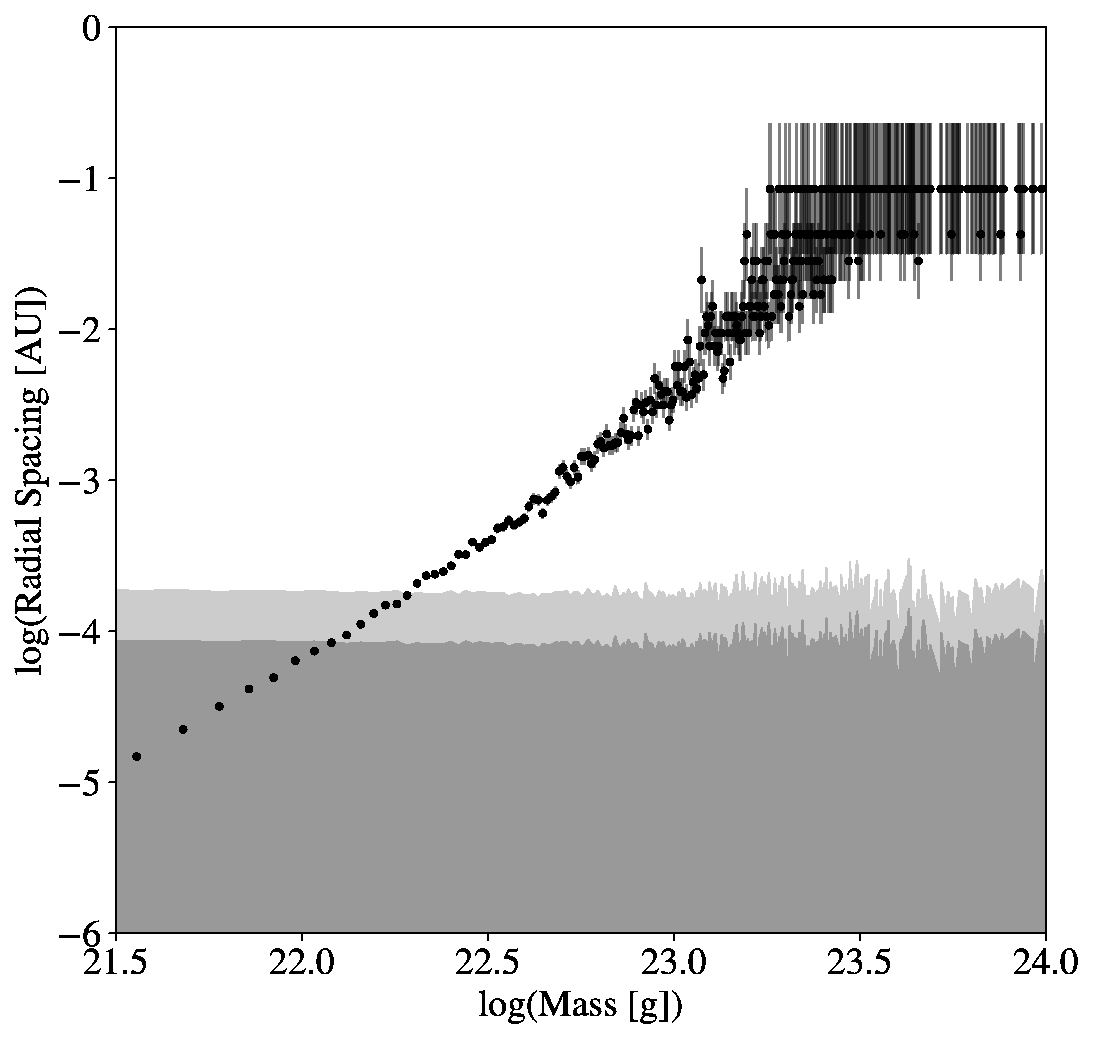}
    \caption{The average spacing in semi-major axis as a function of mass at the end of the high resolution growth simulation (black points). The gray regions indicate the libration width of planetesimals in each mass bin that are in resonance with the most massive oligarch. The light gray region corresponds to the libration width of the 65:64 (highest non-overlapping) resonance and the dark gray region corresponds to the 15:14 (most distant populated) resonance.}
    \label{fig:res_mass}
\end{figure}

\subsection{Resolved Resonances}\label{sec:resonances}

During the oligarchic growth phase, a handful of massive bodies lose energy and angular momentum to the surrounding medium. The previous runaway growth phase leaves behind a steadily varying spectrum of planetesimal masses, whose number decreases with mass. Assuming the planetesimals are randomly distributed throughout the disc, the granularity can be thought of as increasing with mass. This implies that the dynamical friction should be more effective from the smallest planetesimals. Because the planetesimals appear to be more strongly heated below a certain mass threshold, we hypothesize that the change in growth modes has to do with the activation of resonances, rather than a smooth decrease in the stochasticity of scattering events.

Because spiral features like those seen in \citet{2007MNRAS.375..425W, 2007MNRAS.375..460W} and
\citet{2002MNRAS.334...77C} are unlikely to form in such a narrow annulus, we will consider the effects of mean motion resonances (MMRs), which are a subset of the Lindblad resonances considered in the aforementioned studies. In order to determine the resolution required to resolve MMRs, we calculate the libration width 
of first order MMRs associated with the oligarchs and compare it to the average radial spacing between planetesimals as a function mass. The average radial spacing is given by

\begin{equation}\label{eq:spacing}
\left< \Delta r \right> = \frac{\Delta a}{N(m)},
\end{equation}

\noindent where $\Delta a$ is the width of the annulus and $N(m)$ is the number of planetesimals of each mass.

Because the mass distribution has a negative slope and we expect planetesimals of varying mass to be randomly distributed about the disc, the radial spacing between planetesimals should increase with mass. For a fixed resonance width, there should be a cutoff in mass, below which planetesimals are more strongly affected by the resonances. The libration width of a first order mean motion resonance can be derived analytically using the pendulum approximation \citep{2000ssd..book.....M} and is given by

\begin{equation}\label{eq:lib_width}
\frac{\delta a}{a} = \pm \left(\frac{16}{3} \frac{\left| C_{r} \right|}{n} e \right)^{1/2} \left(  1 + \frac{1}{27 j_{2}^2 e^3} \frac{\left| C_{r} \right|}{n} \right)^{1/2} - \frac{2}{9 j_{2} e}  \frac{\left| C_{r} \right|}{n},
\end{equation}

\noindent where $a$ is the semi-major axis at the centre of the resonance, $e$ is the eccentricity of the body in resonance and $j_2 = -q$ where $p:q$ is the MMR being considered. $\left| C_{r} \right|/n = (m^{\prime}/m_{c}) \alpha f_{d}(\alpha)$, where $(m^{\prime}/m_{c})$ is the mass ratio of the body associated with the resonance to the central body, $\alpha$ is the semi-major axis ratio associated with the resonance and $f_{d}(\alpha)$ is the disturbing function. For an interior first order resonance, the disturbing function can be expressed as

\begin{equation}\label{eq:dist}
f_{d}(\alpha) = j b_{1/2}^{j} + \frac{\alpha}{2}\frac{d b_{1/2}^{j}}{d \alpha},
\end{equation}

\noindent \citep{1997A&A...319..290W} where $j = 1 - j_{2}$ and $b_{1/2}^{j}$ is a Laplace coefficient which is defined as

\begin{equation}\label{eq:lap}
b_{s}^{j}(\alpha) = \frac{1}{2 \pi} \int_{0}^{2 \pi} \frac{cos \, j \theta \, d \theta}{\left( 1 - 2 \alpha \, cos \theta + \alpha^2 \right)^{s}}.
\end{equation}

\noindent The derivative in the second term of equation \ref{eq:dist} can be written in terms of the Laplace coefficients \citep{2000ssd..book.....M}

\begin{equation}\label{eq:lap_d}
\frac{d b_{s}^{j}}{d \alpha} = s \left( b_{s+1}^{j-1} - 2 \alpha b_{s+1}^{j} + b_{s+1}^{j+1} \right).
\end{equation}

Figure \ref{fig:res_mass} shows the libration width and average radial spacing of the planetesimals as a function of mass at the end of the high resolution run. To calculate the libration width as a function of planetesimal mass, the $e$ used in equation \ref{eq:lib_width} was taken to be the average eccentricity in each mass bin, while $a$ was set to the semi-major axis of the largest oligarch in the simulation. The slight variation in the resonance width as a function of mass is due to the variation in eccentricity. The error bars on the radial spacing data are calculated from Poisson statistics.

This calculation was done for the 15:14 and 65:64 mean motion resonances. The 15:14 resonance is the most distant first order resonance relative to an oligarch at 1 AU that still lies within the annulus of planetesimals. The 65:64 resonance is the closest MMR for which the libration width is smaller than the spacing between resonances. Higher first order resonances should not exhibit any resolution dependence because they overlap. For planetesimals less massive than about $10^{22}$ g, the average particle spacing drops below the libration width. Bodies below this mass cutoff, which we will refer to as the resonance heating mass $M_{res}$, are more likely to populate MMRs. This provides an explanation for how the oligarchs are preferentially heating the low mass bodies. Comparing with figure \ref{fig:ecc_den_evo}, the mass of the power law break in the surface density matches very closely with $M_{res}$, which also matches the location of the $10^{22}$ g bump in figure \ref{fig:mass_spectrum_evo}. Over the course of the simulation, $M_{res}$ increases by no more than a factor of 2, so the growth mode boundary could easily leave an imprint on the mass spectrum, unlike the shear/dispersion dominated growth boundary, which evolves from $10^{22}$ to $10^{25}$ g.

In the restricted three-body problem, the orbital parameters of a test particle receiving energy and angular momentum via a mean motion resonance will evolve such that \citep{2000ssd..book.....M}

\begin{equation}\label{eq:tiss}
\frac{de}{da} = \frac{a^{3/2} - 1}{2 a^{5/2} e},
\end{equation}

\noindent where $a$ is in units relative to the semi-major axis of the perturbing body. In the above equation, the inclination is assumed to be negligibly small.

Equation \ref{eq:tiss} can be used to place an upper limit on the change in eccentricity that a planetesimal in resonance will experience. Because any $\delta a$ larger than the resonance width will remove the planetesimal from the resonant influence of the oligarch, $\delta e$ is also restricted by the resonance width. Taking $a \approx 0.95$ (the location of the 15:14 resonance), $e \approx 10^{-3}$ and $\delta a \approx 10^{-4}$ (the libration width of the 15:14 MMR) we predict a first order change in eccentricity of $10^{-4}$. This value is small relative to the rms eccentricity of the planetesimals, which explains why the effects of the  mean motion resonances are not visible in figure \ref{fig:ae} or in the top right panel of figure \ref{fig:ecc_den_evo}.

\begin{figure}
    \includegraphics[width=\columnwidth]{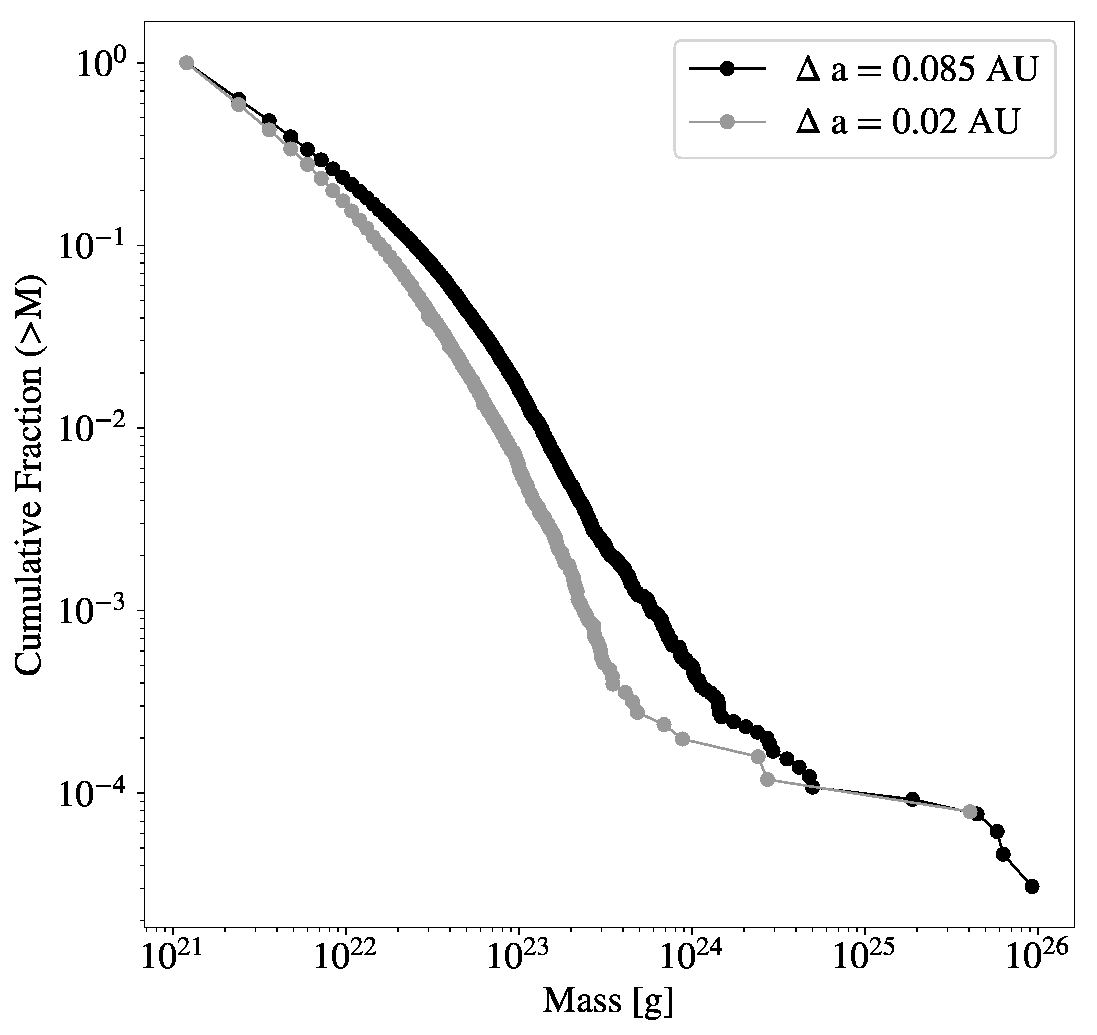}
    \caption{The cumulative fraction of bodies as a function of mass in the wide (solid dark curve) and narrow (gray curve) high resolution growth simulations. Decreasing the annulus width, which depopulates resonances, produces a less prominent $10^{22}$ g bump.
    \label{fig:mass_spectrum_comp}}
\end{figure}

Because the resonances are not possible to pick out by eye on an $a-e$ plot, we ran a modified version of simulation (i), the purpose of which is to show how planetesimal growth proceeds when the mean motion resonances are not present. The initial conditions are identical to those described in section \ref{sec:ics}, except that the annulus only extends from 0.98 to 1.02 AU. This effectively depopulates all of the first order mean motion resonances below the 26:25 resonance. It was not possible to use an annulus skinnier than this without introducing strong boundary effects which influence the growth of the planetesimals. This makes it impossible to depopulate all of the resolved resonances. The mass spectrum at the end of the skinny annulus simulation (also evolved for 20,000 years), along with the mass spectrum at the end of the high resolution growth simulation, is shown in figure \ref{fig:mass_spectrum_comp}. With a narrower annulus, the change in slope of the mass distribution below $10^{22}$ g is noticeably weaker. We attribute this to the fact that the resonant interaction between the planetesimals and oligarchs is diminished because many of the MMRs are now empty.

\subsection{Planetesimal and Oligarch Mixing}\label{sec:mix}

Although figures \ref{fig:res_mass} and \ref{fig:mass_spectrum_comp} provide strong evidence that the $10^{22}$ g bump is being produced by mean motion resonances, it is not immediately clear how the resonances affect such a large fraction of the planetesimals. To have a noticeable effect on the mass distribution, the oligarchs must be affecting bodies below $M_{res}$ everywhere in the disc, not just within the small fraction of space covered by the MMRs at a given time.

We infer that both the oligarchs and planetesimals slowly wander through the annulus due to occasional scattering events. This continually replenishes the population of planetesimals that are sitting inside of resonances. The orbital repulsion effect described by \citet{1998Icar..131..171K}, which moves the oligarchs around, alters the location of the mean motion resonances. In addition, we infer that the planetesimals are occasionally scattered by the oligarchs. These two effects slowly cycle different planetesimals through the resonances and allow the oligarchs to eventually heat a large fraction of the small bodies.

\begin{figure}
    \includegraphics[width=\columnwidth]{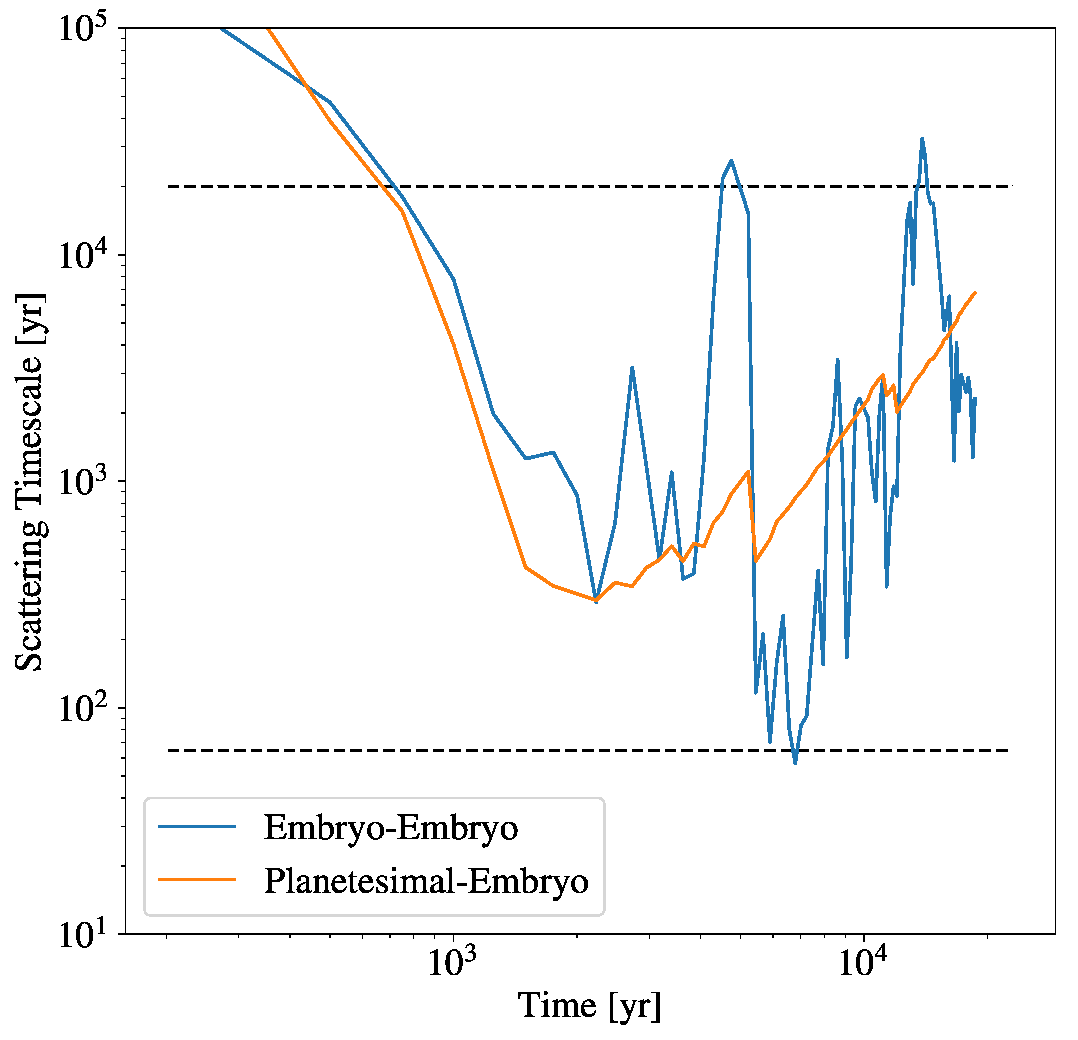}
    \caption{The mean scattering time-scale for embryo-embryo (blue curve) and planetesimal-embryo (orange curve) interactions over the course of the high resolution growth simulation. The dashed lines correspond to the simulation time-scale (upper) and longest relevant mean motion resonance time-scale (lower), which is set by the 65:64 resonance.}
    \label{fig:scatter_timescale}
\end{figure}

In order for this effect to work, the bodies must keep a constant semi-major axis long enough for resonant interactions to play out, while still moving by an appreciable amount over the course of the simulation. The strong scattering time-scale for oligarch-oligarch interactions and oligarch-planetesimal interactions is shown in figure \ref{fig:scatter_timescale}. This is the timescale over which bodies of mass $m$ moving with speed $v$ relative to the local Keplerian velocity will approach each other with an impact parameter $b \leq G m / v^{2}$. This is sufficient to cause a large deflection and effectively randomize the orbital elements of the bodies. In both cases, the average time between scattering events falls between the longest resonance time-scale (set by the 65:64 resonance) and the growth time-scale (set by the duration of the simulation). This indicates that the scattering is vigorous enough to occur many times over the course of the simulation, while still being slow enough to allow mean motion resonances to act. The strong scattering rate is given by

\begin{equation}\label{eq:scatter}
\dot{\overline{N}} \approx \frac{\Sigma}{m} \Omega_{p} R_{h}^2 \left( \frac{v_{h}}{v} \right)^{4},
\end{equation}

\noindent \citep{2006ApJ...651.1194M} where $\Sigma$ and $m$ are the surface density and individual masses of the bodies being scattered, $\Omega_{p}$ is the orbital angular velocity of the bodies and $R_{h}$ and $v_{h}$ are the Hill radius and Hill velocity of the object doing the scattering.

As discussed in section \ref{sec:lowvshigh}, viscous stirring plays an important role in the dynamical evolution of the planetesimals. The effects of viscous stirring are realized over many weak ($b \gg G m / v^{2}$) but frequent encounters. For a population of equal mass planetesimals, the timescale for viscous stirring is given by \citep{1993Icar..106..210I}

\begin{equation}\label{eq:vs_timescale}
    \tau_{vs}  = \frac{\left< e^2 \right>}{d \left< e^2 \right> / dt} \approx \frac{1}{40}\left(\frac{\Omega^{2} a^{3}}{2 G m}\right)^{2} \frac{4 m \langle e^{2} \rangle^{2}}{\Sigma a^{2} \Omega},
\end{equation}

\noindent where $a$ and $e$ are the semi-major axes and eccentricities of the individual planetesimals. By using the properties of the planetesimal disk at the beginning of simulation (i) in equation \ref{eq:vs_timescale}, we find that the viscous stirring timescale is approximately 1000 years. This timescale can be taken as a lower limit because the eccentricity dispersion grows over time.


We also briefly consider the importance of planetesimal-driven migration. This is the coherent change in the semi-major axis of an oligarch due to repeated weak encounters with planetesimals. An upper limit on the planetesimal-driven migration rate of an oligarch is given by \citep{2000ApJ...534..428I}

\begin{equation}\label{eq:mig}
    \left| \frac{d a}{d t} \right| = \frac{a}{T} \frac{4 \pi \Sigma a^{2}}{M_{*}},
\end{equation}

\noindent where $a$ is the semi-major axis of the oligarch, $T$ is the orbital period of the oligarch and $\Sigma$ is the local surface density of the planetesimal disk. For an oligarch on a 1 AU orbit in a planetesimal disk with $\Sigma$ = 10 g cm$^{-2}$, the maximum migration rate is roughly $10^{-5}$ AU / yr. \citet{2009Icar..199..197K} showed that this migration rate is greatly reduced for planetesimals whose encounters are dispersion dominated. In simulation (i), the Hill eccentricity of the smallest planetesimals at the onset of oligarchic growth is larger than 10, in which case the migration rate is reduced by a factor of at least 100 (see figure 7 of \citet{2009Icar..199..197K}). At this rate, it would take an oligarch roughly 1000 years to migrate by a single resonance width ($10^{-4}$ AU). This is much longer than the resonance timescale and therefore we do not expect that planetesimal-driven migration would disrupt the resonant configuration.

From these results, we conclude that the two simultaneous modes of growth are driven by a difference in the way that loosely and densely packed populations of planetesimals exert dynamical friction on large bodies. In the tightly packed case, energy transfer between the oligarchs and planetesimals is facilitated by mean motion resonances with the largest bodies.

\begin{figure*}
    \includegraphics[width=1.0\textwidth]{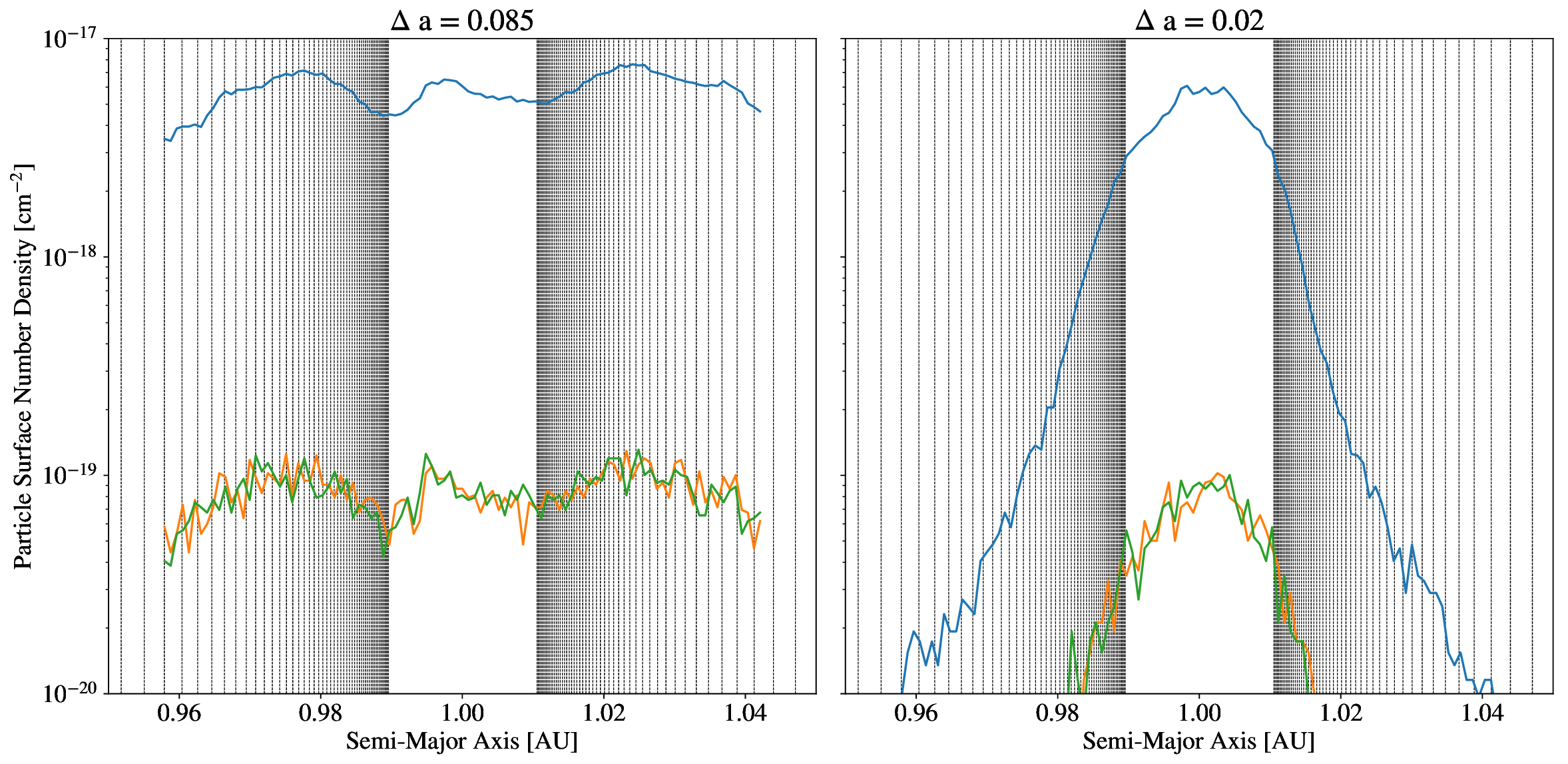}
    \caption{Number density profiles of the planetesimals at the end of the $e_{hi}$ (left) and $e_{narrow}$ simulations (right). The blue curves represent the number density of planetesimals less massive than the resonance heating mass ($2 \times 10^{22}$ g) and the orange curves represent the number density of planetesimals above this mass. The green curves show the surface number density of planetesimals randomly drawn from the low mass bin, such that the total number of bodies matches that of the high mass bin. The vertical lines indicate the positions of first order mean motion resonances with an embryo at 1 AU.}
    \label{fig:num_den}
\end{figure*}

\subsection{Collisionless Dynamics}\label{sec:dynint}

So far, the most compelling evidence of the resonance heating effect that we have shown is the power law break in the surface density distribution in the bottom right panel of figure \ref{fig:ecc_den_evo}. Because the surface density depends on both the mass distribution and spatial distribution of the planetesimals, it is difficult to tell whether the power law break is caused by resonances moving the planetesimals around or is simply set by the mass distribution. To clear up this ambiguity, we ran an additional set of simulations in which a large planetary embryo is embedded in an annulus of planetesimals, this time ignoring the effects of collisions. This forces the mass distribution to be static.

 The initial conditions for these simulations were taken from intermediate snapshots from the runs described in section \ref{sec:results}. Specifically, the initial conditions are taken from the end of the runaway growth phase, which correspond to the snapshots shown in the top row of figure \ref{fig:mass_spectrum_evo}. Additionally, a planetary embryo is placed in the centre of the annulus at 1 AU with an eccentricity of 0.02 and an inclination of 0.01. The mass of the embryo is set at $M = 10^{26}$ g, which is approximately the mass of the largest oligarch at the end of the high resolution simulation. Because collisions are ignored, close encounters are handled with a gravitational softening parameter, the length of which is set to the physical radius of the planetesimals. Both runs are integrated for 2,000 years with fixed timestpng of 0.0025 years. We will refer to the collisionless versions of these simulations as $e_{low}$ and $e_{hi}$, respectively. To further demonstrate that the dynamical excitation of the small planetesimals is driven by mean motion resonances, we also ran a high resolution collisionless simulation with planetesimals outside of the $a$ = 0.99 to 1.01 AU range excluded. We will refer to this simulation as $e_{narrow}$.

As we saw previously, the resonance heating effect manifests itself as an increase in the spacing of the low mass planetesimals. Figure \ref{fig:num_den} shows the planetesimal surface number density at the end of the $e_{hi}$ (left) and $e_{narrow}$ (right) simulations described above. The vertical dashed lines indicate the locations of the non-overlapping first order mean motion resonances with the embryo. The blue curves show the number density of planetesimals below the resonance heating mass, which is $2 \times 10^{22}$ g in this case. The orange curves show the number density of planetesimals above this mass. Finally, to demonstrate that the difference in dynamical behavior between the low and high mass planetesimals is a population effect, the green curve shows the number density of a subsample of planetesimals randomly drawn from the low mass group, such that the total number of planetesimals in the subsample matches that of the high mass group.

In both cases, the particle number density is slightly enhanced around 1 AU, which is due to bodies trapped in the corotation resonance with the embryo. For the wide annulus, the surface number density decreases and then begins to increase again approximately 0.01 AU away from the embryo. As is evident in the left panel of figure \ref{fig:num_den}, the location at which the number density begins to increase corresponds to the location of the closest non-overlapping (j $<$ 65) mean motion resonances. Due to conservation of the Jacobi energy (see equation \ref{eq:tiss}), interior MMRs will push bodies inward, while exterior resonances move bodies outward. This effect was also observed by \citet{2000Icar..143...45R} (see model B) in that the density decreases to the right of interior resonances and increases to the left. Because our setup contains many closely spaced resonances, the cumulative effect is that the density smoothly increases as one moves through the resonances. 

Hence, the number density distribution acquires a 'W' shape around the embryo, with the low points corresponding to the inner edges of the resonant regions. An inspection of the earlier snapshots from this simulation shows that the 'W' structure becomes deeper and narrower with time. This is consistent with our resonance heating argument because the resonance time-scale decreases as one moves away from the embryo. The stronger, closer resonances become effective last, causing the profile to deepen and narrow with time.

The strength of the resonance heating effect depends on the number density of bodies outside of the $0.99 < a < 1.01$ AU region, where the resonances are not overlapping. The 'W' shaped structure relative to the noise of the high mass (orange) and subsampled population (green) is weak compared to the low mass population (blue) due to the fact that there simply aren't many bodies sitting within the resonances. This structure is entirely absent from the narrow annulus, shown in the right hand panel of figure \ref{fig:num_den} because the planetesimal number density near the resonances is too low. We infer that this decrease in number density of the low mass bodies adjacent to the embryo must also be present in the high resolution growth simulation and is enhancing the collision rate below the resonance heating mass.

\begin{figure}
    \includegraphics[width=\columnwidth]{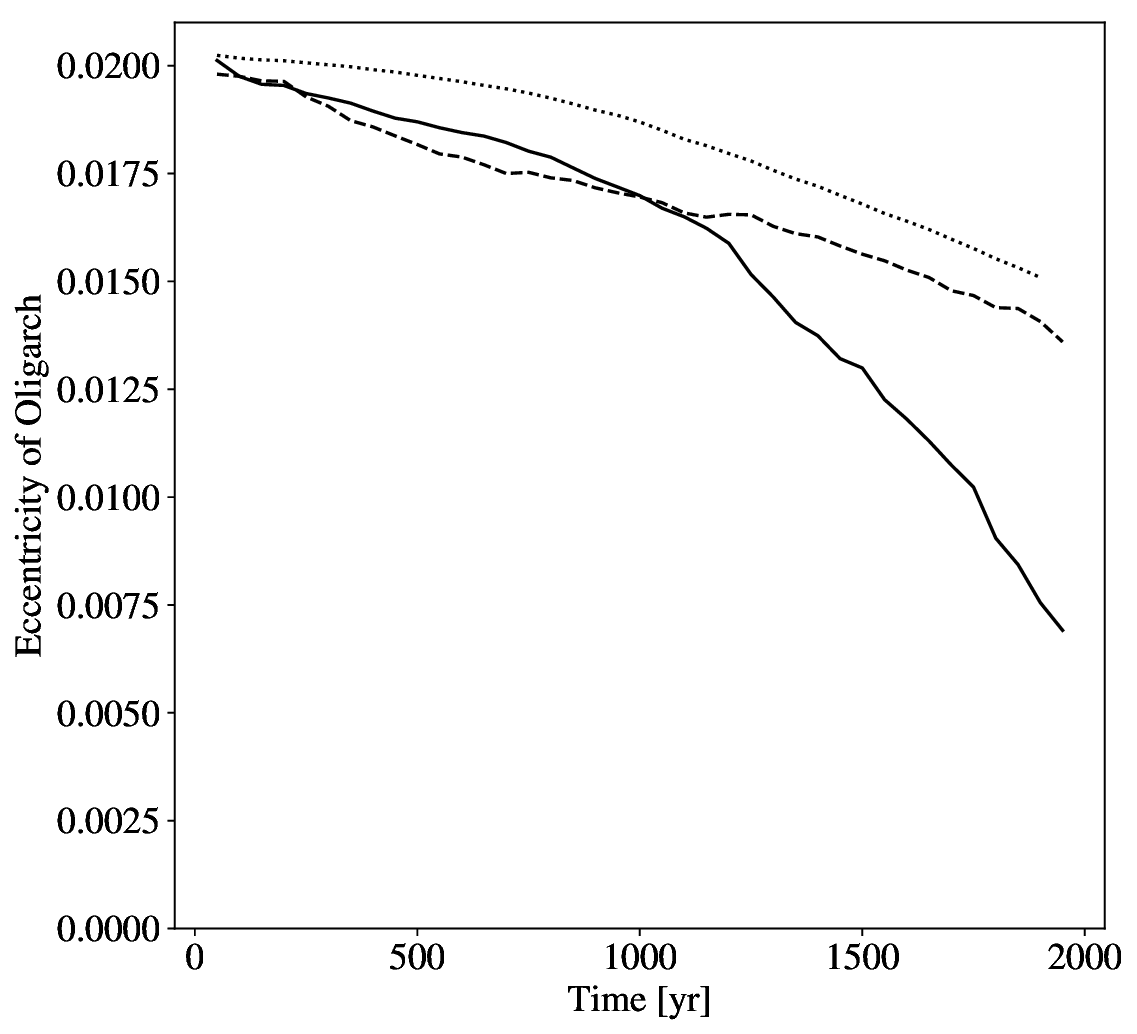}
    \caption{Time evolution of the eccentricity of the oligarch in the $e_{hi}$ (solid), $e_{narrow}$ (dotted) and $e_{low}$ (dashed) simulation.}
    \label{fig:e_oli_evo}
\end{figure}

We also examine the eccentricity evolution of the planetary embryo in the three collisionless simulations, shown in figure \ref{fig:e_oli_evo}. A decrease in the eccentricity of the embryo indicates that energy is being lost to the planetesimals. A steeper drop in eccentricity implies that the exchange happens more quickly and that the effects of dynamical friction are stronger. Only when the resonances are properly resolved and populated does the exchange appear to happen quickly. In both the $e_{narrow}$ and $e_{low}$ simulations, the decrease in eccentricity of the embryo is gradual. This implies that dynamical friction is stronger and energy and angular momentum exchange between the oligarchs and planetesimals proceeds faster when the mean motion resonance heating is effective. In the $e_{hi}$ simulation, the eccentricity of the embryo begins to drop more steeply after about 1000 years. This effect is also noticeable for the $e_{narrow}$ simulation, although it happens sooner. Because we are suddenly dropping a massive body into the annulus of planetesimals, the system likely requires some time to return to a state of quasi-equilibrium. The two-body relaxation time, which is well-described by the viscous stirring timescale \citep{1993Icar..106..210I} appears to roughly coincide with the change in slope for each of the curves in figure \ref{fig:e_oli_evo}, although the connection between the relaxation of the planetesimals and the eccentricity evolution of the oligarch is not immediately clear. In the $e_{low}$ simulation, the change in slope may not be visible due to the fact that the two-body relaxation time is nearly instantaneous due to the small particle count (the relaxation time scales as $N / \ln N$).

To verify the effectiveness of the resonant interactions, we next examine the evolution of the resonant arguments. The libration frequency of a planetesimal in resonance with an embryo is given by \citep{2000ssd..book.....M}

\begin{equation}\label{eq:lib_period}
    \omega_{0}^{2} = -3 j_{2}^{2} C_{r} n e^{\left| j_{4} \right|},
\end{equation}

\noindent where $j_{2}$ = -15 and $j_{4}$ = -1 for the 15:14 MMR and $n = \sqrt{G M_{*} / a^{3}}$ is the mean motion of the planetesimal. For a planetesimal with a typical eccentricity of $10^{-3}$ inside the 15:14 mean motion resonance with a $10^{26}$ g embryo, the libration period is approximately 1000 years. For this reason, it is likely that a particle will undergo only a partial libration cycle before being removed from the resonance by two-body scattering. This makes it difficult to verify the resonant interaction by searching for libration, rather than circulation of the resonant arguments. However, this also implies that the resonances will cause permanent changes to the energy and angular momentum of the planetesimals. In canonical perturbation theory, the action conjugate to the resonant argument should evolve secularly because there is a near-constant term in the partial derivative of the Hamiltonian with respect to the resonant angle. If the viscous stirring timescale, which is the mechanism responsible for the removal of planetesimals from resonance is short compared to the libration period, this secular interaction produces a permanent change in the action (and therefore the energy and angular momentum) of a planetesimal (see the text following equation 10 in \citet{2007MNRAS.375..425W} for a further discussion of this).

\begin{figure}
    \includegraphics[width=\columnwidth]{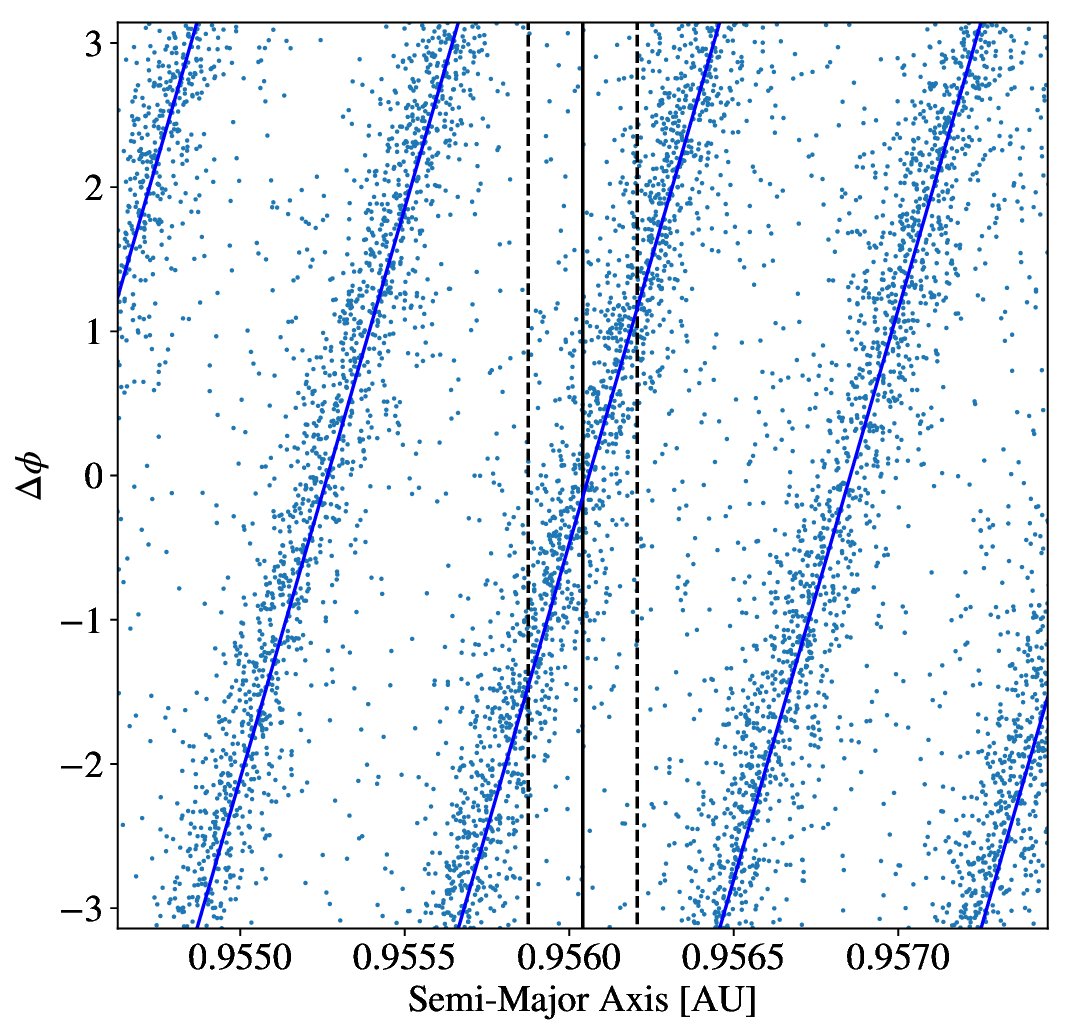}
    \caption{The change in the resonant argument for the oligarch and planetesimals near the 15:14 MMR over a 50 year time interval from the $e_{hi}$ simulation. The solid vertical line designates the nominal resonance location and the dashed vertical lines indicate the minimum libration width. The diagonal blue lines represent the change in $\phi$ due to the Keplerian shear of the disc, which is calculated analytically.}
    \label{fig:lib_arg}
\end{figure}

The typical change in the resonant arguments of the planetesimals from the $e_{hi}$ simulation over multiple synodic periods is shown in figure \ref{fig:lib_arg}. The resonant argument is given by \citep{2000ssd..book.....M}

\begin{equation}\label{eq:res_arg}
    \phi = j_{1} \lambda_{e} + j_{2} \lambda_{p} + j_{4} \varpi_{p},
\end{equation} where $j_{1} = -j_{2} - j_{4} = 16$ for the 15:14 MMR. $\lambda_{e}$ and $\lambda_{p}$ are the mean longitudes of the embryo and planetesimal, respectively and $\varpi_{p}$ is the longitude of pericenter of the planetesimal. The $\Delta \phi$ shown in figure \ref{fig:lib_arg} is the change in this resonant argument between t=1500 and t=1550 years, well after the system has relaxed back to a state of quasi-equilibirum. The blue diagonal lines represent the change in the resonant argument due to the Keplerian shearing of the disc. This quantity is derived by taking the time derivative of equation \ref{eq:res_arg},  which is given by

\begin{equation}\label{eq:res_arg1}
    \frac{d \phi}{dt} = j_{1} \left( \frac{d \varpi_{e}}{dt} + \frac{d M_{e}}{dt} \right) + j_{2} \left( \frac{d \varpi_{p}}{dt} + \frac{d M_{p}}{dt} \right) + j_{4} \frac{d \varpi_{p}}{dt}.
\end{equation} Only the mean anomaly $M = n t$ changes due to the Keplerian shear and so

\begin{equation}
    \Delta \phi_{Kepler} = \sqrt{G M_{*}} \left( j_{1} a_{e}^{-3/2} + j_{2} a_{p}^{-3/2} \right) \Delta t,
\end{equation} where $\Delta t$ is the time interval that we are considering. Note that at the nominal resonance location, $\Delta \phi_{Kepler} = 0$.

Over a 50 year time interval, which is approximately 3 times longer than the synodic period with the embryo, the change in the resonant arguments of the planetesimals appear to be dominated by the differential rotation of the disc. Near the nominal resonance location, the change in $\phi$ is small. The fact that many planetesimals remain within the resonance width and undergo only small changes in $\phi$ over multiple synodic periods demonstrates that the resonant interaction between the planetesimals and the embryo is able to proceed. Because the libration period is not short compared to the scattering timescale, the resonant interactions will cause permanent changes to the energy and angular momentum of the planetesimals.

\section{Implications of Simplifying Assumptions}\label{sec:assumptions}

\subsection{Gas Drag}

Although the effects of gas drag are weak during the planetesimal accretion stage, we will briefly consider its effect to ensure that it would not alter or remove the $10^{22}$ g bump. One way to describe the importance of gas drag on a planetesimal is with the stopping time \citep{1976PThPh..56.1756A}. In the Stokes regime, where the mean free path of gas particles is much smaller than the radius of the planetesimals ($\lambda \ll$ s), the stopping time is given by

\begin{equation}\label{eq:t_stop}
    t_{s} = \frac{2 m}{C_{D} \pi s^{2} \rho_{g} v}.
\end{equation}

\noindent Here, $\rho_{g}$ is the local density of the gas and $C_{D}$ is the drag force coefficient, which is of order unity in this regime. The gas density of the solar nebula is approximately $2 \times 10^{-9}$ g cm$^{-3}$ at 1 AU \citep{1981PThPS..70...35H}.

The relative velocity between the planetesimals and the gas is set by both the random motions of the planetesimals and the fact that the gas orbits at a sub-Keplerian speed due to its internal pressure support. The planetesimal velocity due to random motions is given by \citep{1993prpl.conf.1061L}

\begin{equation}\label{eq:vrnd}
    v_{rnd} = v_{k} \sqrt{\left< e^2\right> + \left< i^2\right>}.
\end{equation}

\noindent At the beginning of the simulation (i), $v_{rnd}$ is on the order of $10^3$ cm s$^{-1}$. The headwind speed due to the sub-Keplerian motion of the gas is given by \citep{1976PThPh..56.1756A}

\begin{equation}\label{eq:vgas}
    v_{gas} = v_{k} \left( 1 - (1 - 2 \eta)^{1/2} \right).
\end{equation}

\noindent At 1 AU in the solar nebula, $\eta \approx$ 0.002, which gives $v_{gas} \approx$ 6000 cm s$^{-1}$. From equation \ref{eq:t_stop}, the smallest planetesimals in simulation (i) have a stopping time on the order of $10^4$ years. This is much longer than the synodic period of the relevant planetesimal-oligarch resonances, therefore we expect the dynamics of even the smallest planetesimals to be entirely dominated by the resonances.

Additionally, gas drag can drive the inward radial drift of small bodies. Although the relatively large stopping time of the planetesimals suggests that this effect should be unimportant, the resonances are narrow enough that even a small amount of radial drift could quickly remove planetesimals from their influence. The radial drift velocity of a particle is given by \citep{1977MNRAS.180...57W}

\begin{equation}\label{eq:v_drift}
    v_{r}  = -\frac{2 v_{gas}}{\Omega t_{s} + \left( \Omega t_{s} \right)^{-1}}.
\end{equation}

\noindent where $\Omega$ is the orbital angular frequency of the body being considered. Using the values for $v_{gas}$ and $t_{s}$ described above, the inward radial drift speed of the smallest planetesimals is 0.1 cm s$^{-1}$. At this rate, a planetesimal will drift across a typical resonance width ($10^{-4}$ AU) in about 1000 years. This is comparable to the viscous stirring timescale, which suggests that radial drift should not have a significant effect on the resonant dynamics.

Massive bodies can create density waves in the gas disk which carry away angular momentum \citep{1979ApJ...233..857G, 1980ApJ...241..425G} through an effect known as type I migration (for a recent review see \citet{2014prpl.conf..667B}). The strength of this effect scales linearly with the mass of the body and should therefore be most effective for planetary embryos. In addition to gas drag causing planetesimals to radially drift across resonances, type I migration can cause planetary embryos to drift inwards, potentially moving the resonances before they have a chance to act on the planetesimals. \citet{2002ApJ...565.1257T} provides an analytic expression for the torque exerted on a body embedded in a three-dimensional gaseous isothermal disk as

\begin{equation}\label{eq:t1mig_torque}
     \Gamma = -\left( 1.364 + 0.541 \alpha \right) \left (\frac{m}{M_{*}} \right)^{2} \left( \frac{r \Omega}{c_{s}} \right)^{2} \Sigma r^{4} \Omega^{2},
\end{equation}

\noindent where $\alpha$ is the power law index of the radial surface mass density profile $\Sigma \propto r^{-\alpha}$ of the gas and $c_{s}$ is the local sound speed of the gas, $m$ is the mass of the body, $\Omega$ is the orbital angular frequency of the body and $r$ is the orbital distance of the body. The radial migration velocity is given by

\begin{equation}\label{eq:t1mig_rate}
    \dot{r} = \frac{-2 r \Gamma}{L},
\end{equation}

\noindent where $L = m \left( G M_{*} r \right)^{1/2}$ is the total angular momentum of the body. For a $10^{26}$ g embryo orbiting at 1 AU and taking $\alpha$ = 3/2, $\Sigma$ = 1700 g cm$^{-2}$ and $c_{s}$ = $10^{5}$ cm s$^{-1}$ \citep{1981PThPS..70...35H}, the inward migration speed is $2 \times 10^{-7}$ AU yr$^{-1}$. At this rate, the 15:14 MMR with an embryo will migrate by a entire resonance width in roughly 500 years, which is much longer than the synodic period. It should be noted that the analysis of \citep{2002ApJ...565.1257T} has been shown to produce migration rates that are up to a factor of $10^{3}$ too large to be consistent with observed populations of exoplanets \citep{2004ApJ...604..388I, 2005A&A...434..343A, 2011MNRAS.417..314M}. The migration timescale derived above should therefore be taken strictly as a lower limit.

As migration proceeds, planetesimals can have their spatial distribution altered as they become trapped in exterior resonances with protoplanets \citep{1985Icar...62...16W}. A necessary condition for resonant trapping to occur is that the drift timescale of a planetesimal across a resonance must be much longer than the libration period \citep{1988Icar...76..295D}. As we calculated in section \ref{sec:dynint}, a typical libration period is around 1000 years. This is comparable to the timescale for \textbf{two-body} scattering and also for radial drift of planetesimals across a resonance due to gas drag. Therefore, we do not expect that exterior resonances should be effective at trapping drifting planetesimals.

\subsection{Inflated Collision Cross Section}

As discussed in section \ref{sec:collModel}, enhancing the geometric collision cross section $f$ of the planetesimals by a factor of 6 only slightly reduces the effectiveness of gravitational scattering. More importantly, varying $f$ alters the accretion timescale. The implication of this is that a longer accretion timescale causes oligarchic growth to commence later when the disk is more dynamically excited. The dynamical excitation of the disk, which is set by the rms eccentricity, alters the libration width of planetesimals in resonance with the oligarchs. In section \ref{sec:resonances}, we argued that the intersection between the libration width and the radial spacing between planetesimals sets the location of the bump in the mass distribution. Because the radial spacing between planetesimals increases with mass and the libration width of the resonances gets larger with time, the mass of the bump would be larger had we used $f$ = 1.

It is difficult to predict exactly how much $f$ reduces the accretion timescale, but \citet{1996Icar..123..180K} showed that it is by no more than factor of $f^2$. Integrating equation \ref{eq:vs_timescale}, the rms eccentricity scales with $t^{1/4}$. Having enhanced the collision cross section by a factor of 6, the rms eccentricity should be a factor of $(f^2)^{1/4} \approx 2.5$ times larger with a realistic collision cross section. Using equation \ref{eq:lib_width}, a factor of 2.5 increase in the eccentricity of a planetesimal increases the libration width by a factor of about 1.6. Looking at figure \ref{fig:res_mass}, this would cause an extremely insignificant change in the resonance heating mass.

\subsection{Fragmentation}

Because our model does not include the effects of collisional fragmentation, we examine the statistics of the collisions in simulation (i) to estimate its effects had it been included. Here, we find that roughly 60 percent of the collisions occur with a relative velocity larger than the mutual escape velocity of the two planetesimals. Assuming that the planetesimals are rubble piles with no internal strength, these high velocity collisions should be completely disruptive. Previous studies of planetesimal accretion have shown that including the effects of fragmentation tends to slow down the accretion process, but does not qualitatively change it \citep{1993Icar..106..190W, 2005ApJ...625..427L}.

If we separate the collision statistics into those occurring between bodies below the bump mass ($< 10^{22}$ g) and those above the bump mass, we do not see a significant difference in the relative amount of catastrophic collisions. Above the bump mass, about 60 percent of collisions occur at disruptive velocities, while 70 percent of collisions between bodies below the bump mass occur at these high relative velocities. Because we do not have multiple simulations to draw from, it is difficult to estimate errors to tell whether this difference is statistically significant.

Assuming that disruptive collisions are not significantly more common below the bump mass, we expect that fragmentation would alter our results by lengthening the accretion timescale. As discussed in the previous section, the location of the bump depends on the rms eccentricity of the planetesimals during the oligarchic growth phase. If this phase takes longer to commence, the libration widths of the resonances would be larger when the embryos form, which increases the mass below which the resonant heating is effective. Although it is difficult to determine by exactly how much fragmentation extends the accretion timescale, statistical models of planetesimal growth show that $10^{26}$ g embryos can be formed in $10^{5}$ years \citep{1993Icar..106..190W} when the effects of fragmentation are included. We showed in the previous section that a factor of 36 increase in the accretion timescale (720,000 years) makes a negligible difference in the location of the bump in the mass spectrum. Similarly, we expect that including the effects of fragmentation would make a negligible contribution to the bump mass.

\section{Dependence on Initial Planetesimal Mass}\label{sec:intermed}

In order to demonstrate that our results are robust, and also to gain some insight into how the resonance heating mass is related to the initial planetesimal mass, we ran two more simulations of planetesimal growth at intermediate resolutions. Our choice of initial planetesimal mass in the high resolution simulation was somewhat arbitrary and understanding how this parameter affects the resulting distribution of masses is necessary in order to connect our results with observations of small Solar System bodies.

The configurations used were identical to the high resolution run described in section \ref{sec:ics}, except that the particle count was changed. These two additional runs contained 250,000 and 500,000 starting planetesimals, which corresponds to $m = 2.4 \times 10^{21}$ and $m = 4.8 \times 10^{21}$ g, respectively, at the same surface density. To match the same eccentricity dispersion as the models described in section \ref{sec:ics}, we used $\langle e^2 \rangle^{1/2} = 3.17 h/a$ and $\langle e^2 \rangle^{1/2} = 2.52 h/a$, with the inclination dispersion set to half of those values.

\begin{figure}
    \includegraphics[width=\columnwidth]{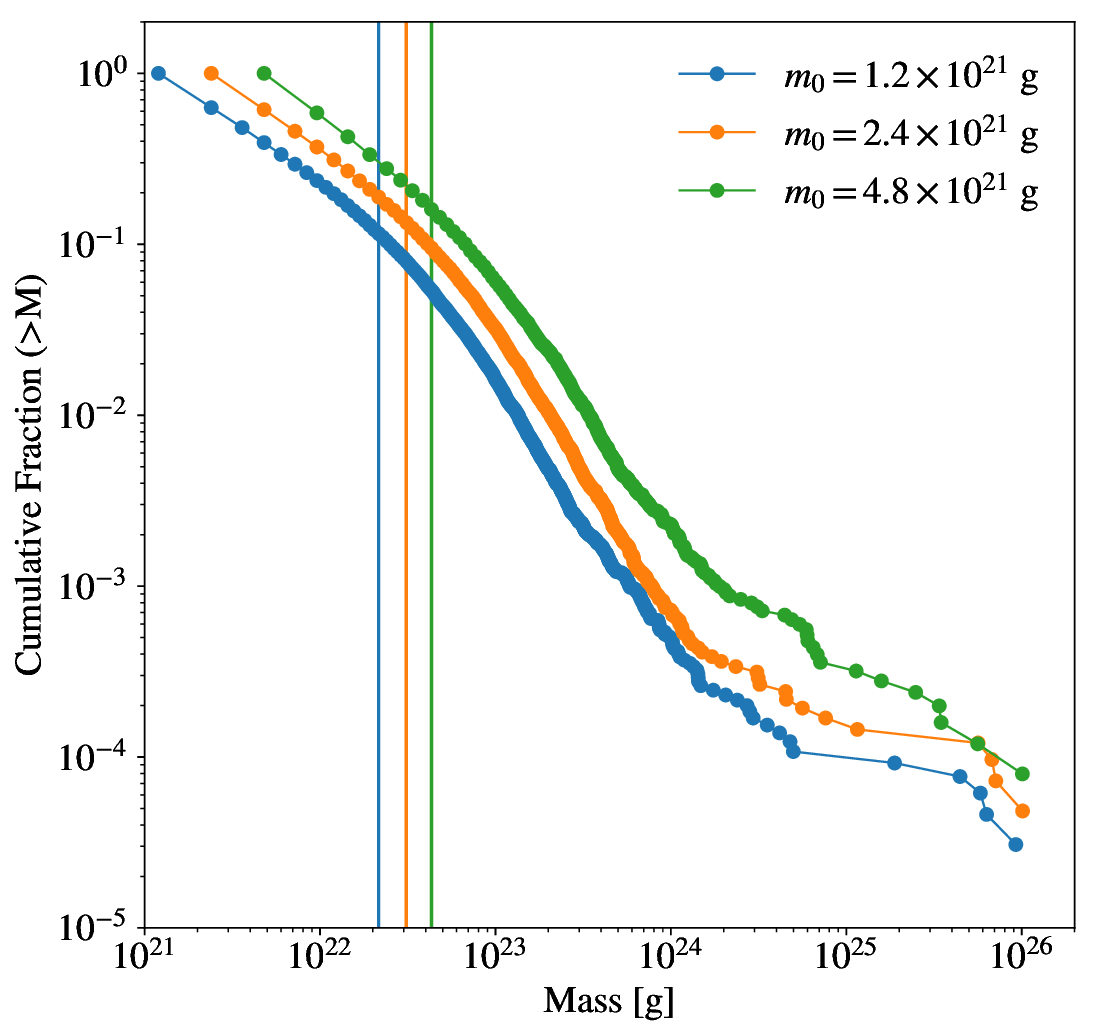}
    \caption{The cumulative mass distribution of planetesimals at the end of the three high-resolution growth simulations. The vertical lines represent the mass at which the planetesimal spacing in semi major axis falls below the libration width of the MMRs.}
    \label{fig:resolution_mass_dist}
\end{figure}

Figure \ref{fig:resolution_mass_dist} shows the mass spectrum at T = 20,000 years in the high resolution growth simulation, along with the intermediate resolution growth simulations mentioned above. In all three cases, the bump in the mass distribution approximately matches up with the resonance heating mass. There is a clear positive trend between the resonance heating mass and the initial planetesimal mass $m$, which we attribute to two things. First, larger planetesimals will be spaced further apart for a fixed disc surface density. This pushes the intersection point between the planetesimal spacing and libration width to higher mass. Secondly, larger initial planetesimals will cause oligarchic growth to begin at a higher mass \citep{2017Icar..281..459M}. This increases the libration width of the resonances, which moves the intersection between planetesimal spacing and resonance width to a higher mass.

\begin{figure}
    \includegraphics[width=\columnwidth]{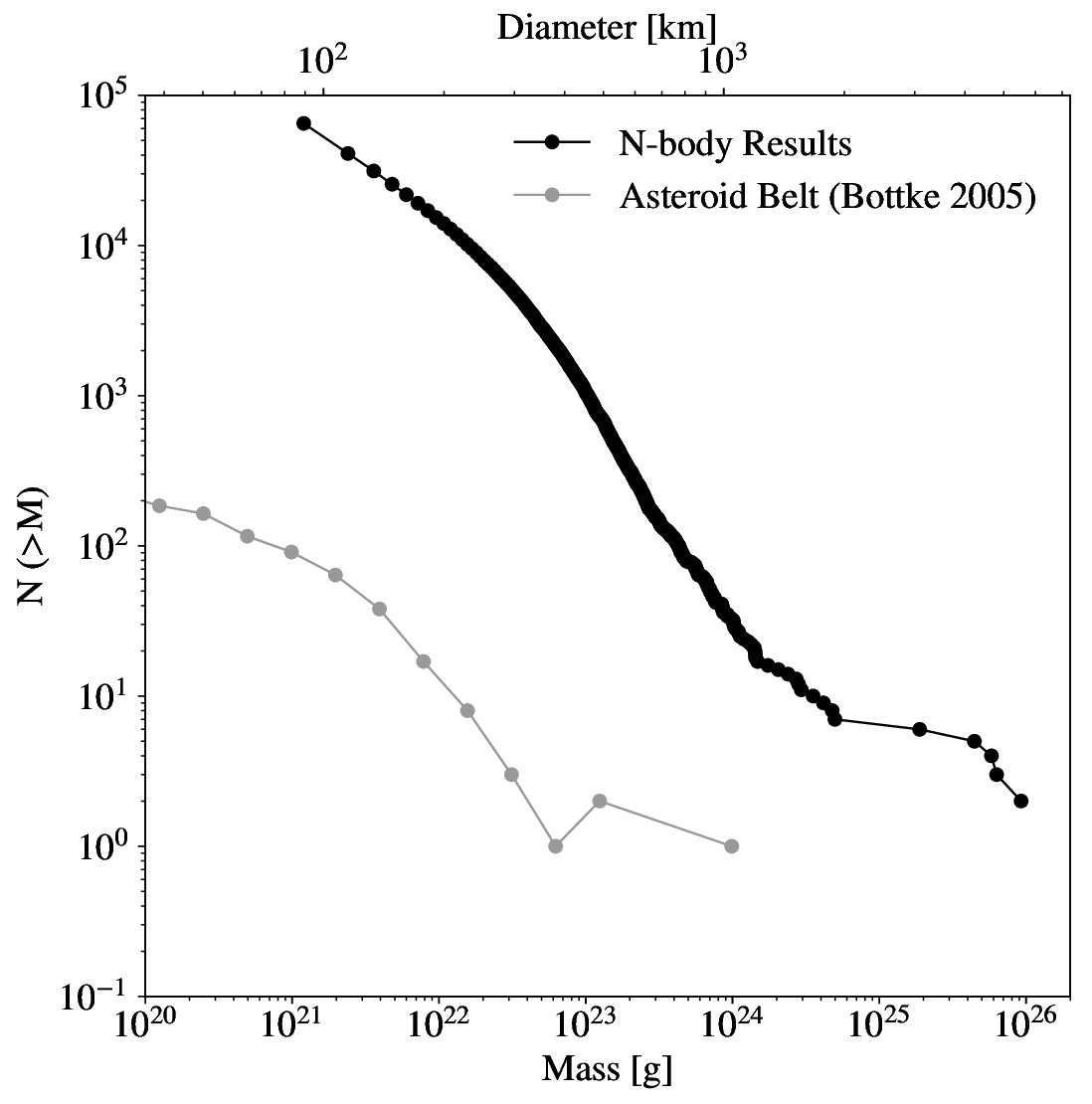}
    \caption{A comparison between the cumulative number of objects in our high resolution growth simulation (black curve) and the present day asteroid belt \citep{2005Icar..175..111B} (gray curve) as a function of size and mass.}
    \label{fig:mass_dist_ast_compare}
\end{figure}

The size frequency distribution of asteroid belt objects is known to exhibit a power law break around around $1.05 \times 10^{21}$ g \citep{2002aste.book...71J}. The fact that the knee in the mass distribution is reproducible and appears sensitive to $m_{0}$ demonstrates that this value could be tuned to constrain planetesimal formation models by matching the observed mass of the bump with simulations. There is evidence that the shape of the SFD for asteroid belt objects more massive than about 100 km reflects that of the primordial population \citep{2009Icar..204..558M}. Although the purpose of this paper is not to try to tune the model to match observations, we show the mass distribution from the $N=10^6$ growth simulation alongside the SFD of asteroid belt objects \citep{2005Icar..175..111B} in figure \ref{fig:mass_dist_ast_compare} for comparison. Although the bump location in all of our simulations is larger than 100 km, a smaller initial planetesimal mass would probably produce a better match. The slope of the mass distribution above the bump mass matches reasonably well in both cases, but the slopes on the low mass end are discrepant. As the results from figure \ref{fig:mass_spectrum_comp} show, a wider annulus tends to produce a shallower slope below the resonance heating mass. A better match might be made by simulating a wider annulus which includes resonances further from the embryos. Unfortunately, simulating an annulus wide enough to resolve all of the first order MMRs would require an extremely large number of particles.

As discussed in section \ref{sec:assumptions}, we do not expect the simplifying assumptions of perfect accretion and the absence of gas drag, along with the artificially inflated collision cross section to have a significant effect on the size distribution of accreted bodies. Because we have proposed that the initial planetesimal size could be reduced to better match the break in the asteroid belt SFD, we briefly consider the smallest sized planetesimals for which the resonant heating effect is not damped by gas drag. This will occur when the synodic period of the resonance becomes comparable to the stopping time. The latter quantity is given by equation \ref{eq:t_stop} and scales linearly with planetesimal size. For 100 km planetesimals, we showed that the stopping time is around $10^4$ years. The stopping time becomes comparable to the synodic period for bodies smaller than a few km in size, at which point our calculations would become invalid.

\section{Summary and Discussion} \label{sec:discussion}

We have revealed a new mode of growth during the planetesimal accretion phase by simulating the interaction between oligarchs and planetesimals at unprecedented resolution. Shortly after the onset of oligarchic growth, a bump develops in the mass distribution of planetesimals. Below the bump mass, the surface density of planetesimals follows a shallower power law distribution. The break occurs near the mass at which the radial spacing of planetesimal matches the libration width of first order mean motion resonances with the oligarchs. These resonances, which are preferentially populated by the more numerous low mass planetesimals, act as effective pathways for dynamical friction to transfer energy and angular momentum from the oligarchs to the planetesimals. This result is analogous to the resolution dependence that \citet{2007MNRAS.375..425W, 2007MNRAS.375..460W} found when examining the interaction between the bar and halo of a galaxy. This also matches the results of \citet{2006Icar..184...39O} and \citet{2002MNRAS.334...77C}, which showed that finer granularity in a planetesimal disc increases the effectiveness of dynamical friction. Bodies below the bump mass, which are packed tightly enough together to populate the resonances receive a disproportionate amount of energy and angular momentum from the oligarchs. Because this happens when the disc is hot enough to render gravitational focusing ineffective, this enhances the growth of the smallest planetesimals and produces a bend in the mass distribution.

Additionally, we ran a high resolution planetesimal growth simulation in which the width of the annulus was limited to exclude some of the resonances. Doing so mostly suppressed the formation of the bump near the resonance heating mass, although it did not completely get rid of it. We attribute this to the fact that we cannot make the annulus narrow enough to exclude all of the important resonances without introducing strong boundary effects which interfere with planetesimal growth. The fact that the feature in the mass distribution was greatly diminished when the many of the resonances were excluded is strong evidence that the MMRs are responsible for creating this bump.

We confirmed this dynamical effect by placing a massive oligarch into a smoothly varying distribution of planetesimal masses. Within a few thousand orbits, the low mass planetesimals sitting within the zone of influence of the resonances began to migrate outwards, leaving a dearth of low mass bodies around the embryo. This effect did not appear when we simulated a similar heterogeneous distribution of masses, but limited the width of the annulus to exclude many of the MMRs. Additionally, the eccentricity of the oligarch decreased much more quickly when placed in a finely resolved disc with populated MMRs.

This result is significant because it suggests a new potentially observable link between the planetesimal formation process and the residual population of planetesimals in the present day Solar System. We showed that the mass distribution in our highest resolution simulation looks qualitatively similar to the SFD of objects in the asteroid belt, which, for objects larger than $\approx$ 100 km, reflects the population of planetesimals at the end of the accretion stage \citep{2009Icar..204..558M}. We demonstrated that tuning the initial planetesimal mass in our simulation changes the location of the bump. This could potentially be matched with the observed 100 km feature in the asteroid belt to constrain planetesimal formation models. Additionally, using a wider annulus, which populates more of the first order MMRs, tends to produce a shallower slope on the low mass end of the mass distribution, which matches more closely with observations.

As discussed in section \ref{sec:bump}, there are a couple of explanations for this bump feature which were obtained from statistical models of planetesimal coagulation. It is unlikely that this resonance heating effect would naturally emerge from a statistical growth model unless it was explicitly built in. For this reason, this resonance heating effect has likely not been considered before. To our knowledge, no one has ever run an N-body simulation of planetesimal accretion to the oligarchic growth phase at this resolution.

Our results show that a feature similar to the power law break in the size distribution of asteroid belt objects can be produced without the effects of fragmentation, in contrast to \citet{2009Icar..204..558M}. Although our model does not account for the effects of fragmentation, the statistics of collisions above and below the bump mass are quite similar. We do not expect that including the effects of collisional fragmentation would significantly affect the location or strength of the bump. Regardless, these results demonstrate that a careful treatment of the dynamics is necessary to properly model planetesimal accretion during the oligarchic growth phase.

\section*{Acknowledgements}
We would like to thank the referee Eiichiro Kokubo for his valuable comments. We would also like to thank Rory Barnes for a careful reading of an early version of this manuscript and Derek Richardson for his advice and assistance while we worked to implement the collision detection module in {\sc ChaNGa}. Additionally, we would like to thank, Mario Juric, David Fleming, Daniel Fabrycky and Jake Vanderplas for their helpful feedback and advice throughout this project. This work was facilitated through the use of advanced computational, storage and networking infrastructure provided by the Hyak supercomputer system at the University of Washington. Some of the simulations used in this work were run on XSEDE/TACC Stampede and Stampede2 under allocation TG-MCA94P018. We made use of {\sc PYNBODY} \citep{2013ascl.soft05002P} in our analysis of this paper. SCW and TRQ were partially supported by NSF award ACI-1550234.

\bibliographystyle{mnras}
\bibliography{references}

\section{Appendix A: Mean Motion Resonance Test}

To demonstrate that {\sc ChaNGa} can properly track the motions of bodies in a mean motion resonance over many orbits, we present a test case with a central star, a perturbing massive body and a planetesimal. Although the simulations presented in this paper are far more complex than this three body setup, the resonant interactions that produce the bump in the mass distribution are driven by the most massive bodies in the simulation. Because there are only a handful of these massive bodies, the tree approximation should have a negligible effect on the force contribution from these objects. For this reason, the behavior shown here should also apply to the previously presented simulations of planetesimal growth.

To test that the resonant interaction evolves correctly, we follow the evolution of the planetesimal in the complex x-y plane defined by the variable \citep{1989Icar...82..402D}

\begin{equation}\label{eq:complex}
    z = \left(G M_{*} a\right)^{1 / 4}\left\{2\left[1-\left(1-e^{2}\right)^{1 / 2}\right]\right\}^{1 / 2} \exp \left[i(\overline{\omega}-\lambda)\right],
\end{equation}

\noindent where a, e, $\overline{\omega}$ and $\lambda$ are the semi-major axis, eccentricity, longitude of perihelion and mean longitude of the planetesimal. If these values are recorded at the moment of opposition between the perturber and the planetesimal ($\lambda - \lambda_{p} = \pi$), the trajectory of the planetesimal should trace out a closed loop in the x-y plane. This indicates that an approximate Hamiltonian of the system is preserved and the resonant interaction is properly accounted for.

\begin{figure}
    \includegraphics[width=\columnwidth]{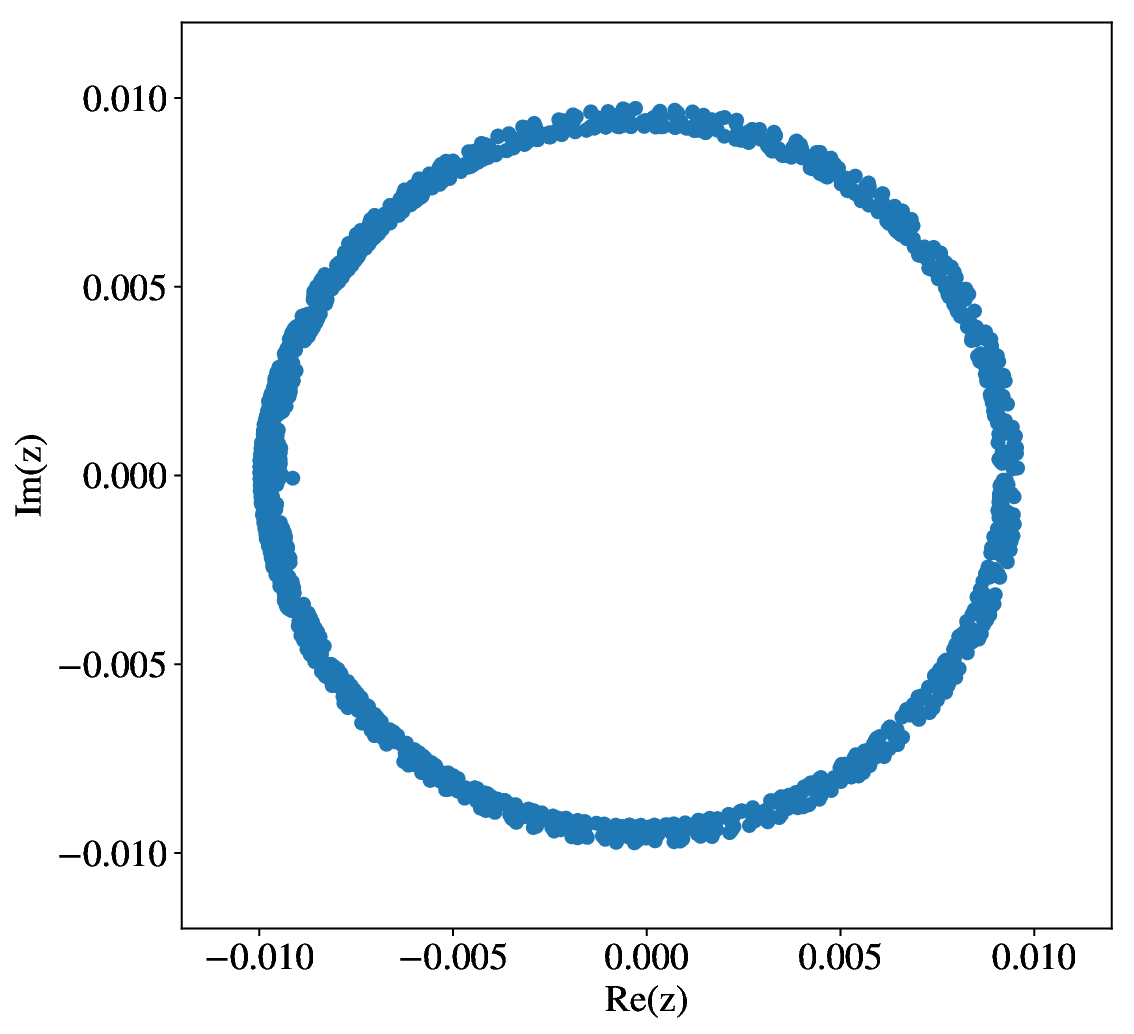}
    \caption{The evolution of the planetesimal in the complex plane defined by equation \ref{eq:complex}. Each point represents the state of the system when the perturber and planetesimal are at opposition.}
    \label{fig:threebody}
\end{figure}

Figure \ref{fig:threebody} shows the evolution of the planetesimal over 20,000 years (about 130 synodic periods) in the complex x-y plane. The orbital elements of the planetesimal at the exact moment of opposition are calculated via linear interpolation. The perturbing body is placed at 1 AU on a circular orbit and has a mass of $10^{26}$ g, which is similar to the final mass of the embryos presented in simulation (i). The planetesimal is given a mass of $1.2 \times 10^{22}$ g and is placed on a coplanar orbit with a semi-major axis of 0.95501, which corresponds to the nominal location of the 15:14 mean motion resonance with the perturber. To start the planetesimal in a stable equilibirum, its orbital eccentricity is set to $10^{-2}$ so that the longitude of pericenter is well-defined. Fixed timestpng of $\Delta$T = 0.0025 yrs are taken, which is the same as the previous simulations. In this configuration, the Jacobi constant is small and so the planetesimal follows a circular trajectory in the complex plane. Because the perturbing body is on a circular orbit, the forced eccentricity is zero and the trajectory of the planetesimal is centered at the origin. Most importantly, trajectory of the planetesimal appears to follow a closed loop, which indicates that an approximate Hamiltonian is preserved and the integration is accurate enough to follow mean motion resonances in this configuration.

\begin{figure*}
    \includegraphics[width=\textwidth]{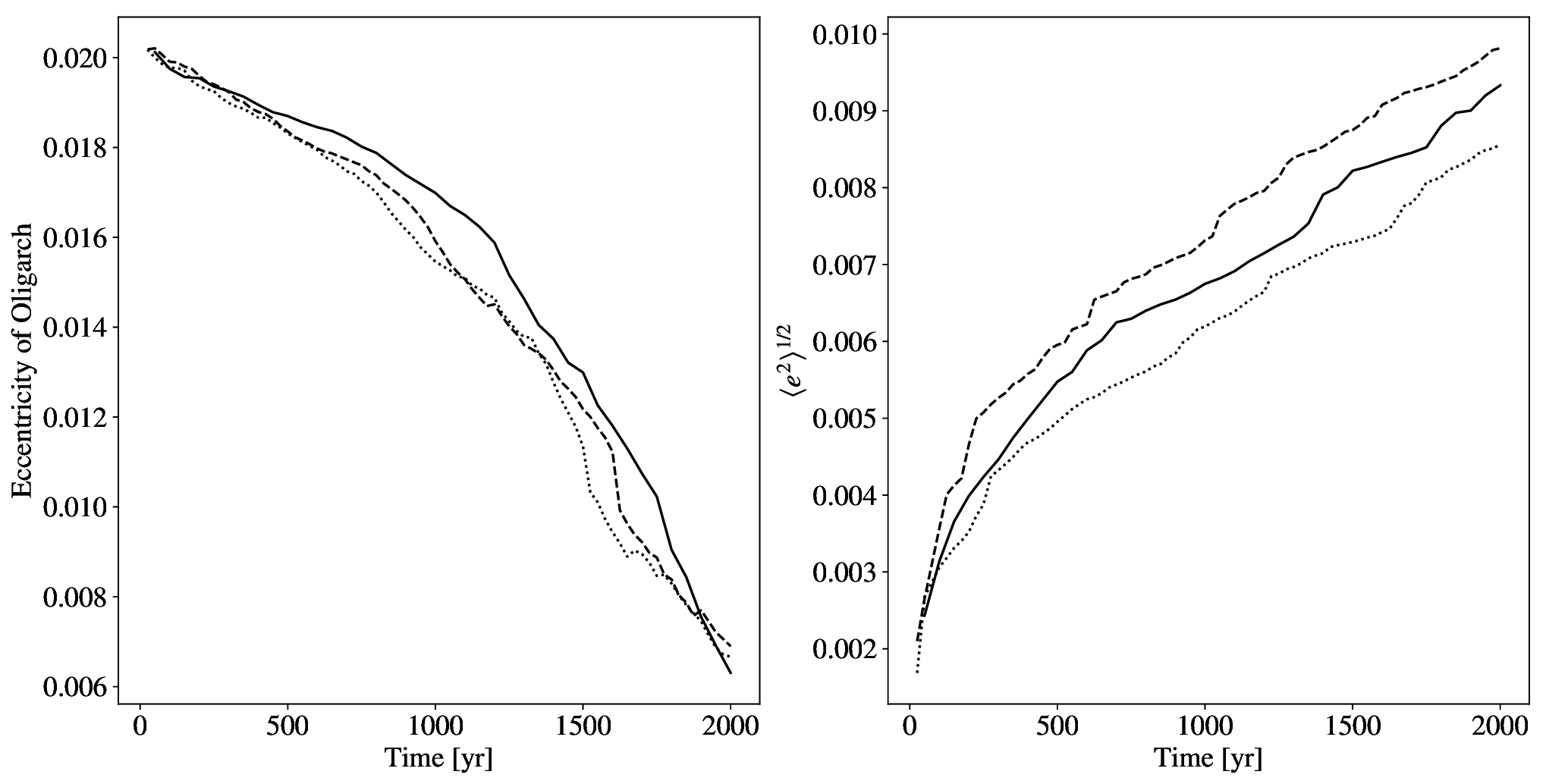}
    \caption{Time evolution of the eccentricity of the oligarch (left) and rms eccentricity of the smallest planetesimals (right) in the $e_{hi}$ simulation with an opening angle $\Theta_{BH}$ of 0.7 (solid line) and 0.35 (dashed line). The dotted line shows $\Theta_{BH}$ = 0.7 with a timestep that is half as large.}
    \label{fig:theta_test}
\end{figure*}

\section{Appendix B: Tree Approximation and Opening Angle}

The results we have presented in this work depend on repeated two-body interactions between planetesimals. As discussed in section \ref{sec:assumptions}, the location of the bump in the mass spectrum depends on the amount of viscous stirring that has occurred before oligarchic growth commences. Additionally, the resonant heating effect presented in this work requires that viscous stirring is not too vigorous as to prevent repeated conjunctions. Here, we examine the impact of force calculation errors from our tree algorithm on these phenomena.

{\sc ChaNGa} calculates the gravitational interaction force between particles via a tree approximation. Particles that only weakly contribute to the gravitational potential are grouped together during the force calculation phase. The opening angle $\Theta$ controls how likely particles are to get grouped together during this stage. Because the lowest mass particles contribute weakly to the surrounding potential, the gravitational contribution from these bodies is more approximate. This can potentially alter the effectiveness of viscous stirring. For this reason, we re-ran the $e_{hi}$ simulation with a smaller, more restrictive opening angle of $\Theta_{BH}$ = 0.35 to test whether the tree approximation is noticeably altering the behavior of viscous stirring. Because we expand forces from tree nodes to hexadecapole order, an opening angle that is a factor of 2 smaller should reduce the error in the force calculations by a factor of 16. Additionally, we test $\Theta_{BH} = 0.7$ with a timestep of $\Delta T$ = 0.00125 years.

A comparison between the three versions of the $e_{hi}$ simulation is shown in figure \ref{fig:theta_test}. In all cases, the oligarch slowly loses eccentricity for about 1000 years before the curve drops more steeply. With a smaller opening angle or a smaller timestep size, the downturn appears to happen slightly sooner. By the end of the simulation, the resulting eccentricities of the oligarchs are still within 10 percent of each other. The right hand panel of figure \ref{fig:theta_test} shows the evolution of the rms eccentricity of the lowest mass planetesimals. There is some divergence between the curves early on, but this only lasts for a fraction of the viscous stirring timescale. As was shown in section \ref{sec:assumptions}, it would take a much larger difference in the rms eccentricity than is seen here here to alter the bump in the mass spectrum. The evolution of the rms inclination of the planetesimals is qualitatively similar in all three cases.

\bsp    
\label{lastpage}
\end{document}